\documentclass[aps,pra,twocolumn,groupedaddress]{revtex4-1}
\usepackage[latin9]{inputenc}
\usepackage{verbatim}
\usepackage{graphicx}
\usepackage{xargs}[2008/03/08]
\usepackage{times}
\usepackage{amsmath}

\newcommand{\modulus}[1]{\left|#1\right|}
\def\w{\omega}
\def\W{\Omega}
\def\ud{\textrm{d}}
\newcommand{\bra}[1]{\left\langle{#1}\right|}
\newcommand{\ket}[1]{\left|{#1}\right\rangle}

\begin{document}

\preprint{}

%\title{Characterising the single-photon spectral-temporal wave function using electro-optic spectral shearing interferometry}

\title{Experimental single-photon pulse characterization by electro-optic shearing interferometry}


\author{Alex O. C. Davis$^{1}$}
\author{Val\'{e}rian Thiel$^{1}$}
\author{Micha\l{} Karpi\'{n}ski{$^{1,2,*}$}}
\author{Brian J. Smith{$^{1,3}$}}
%\thanks{}
\affiliation{{$^{1}$}Clarendon Laboratory, University of Oxford, Parks Road, Oxford, OX1 3PU, UK}
\affiliation{{$^{2}$}Faculty of Physics, University of Warsaw, Pasteura 5, 02-093 Warszawa, Poland}
\affiliation{{$^{3}$}Department of Physics and Oregon Center for Optical, Molecular, and Quantum Science, University of Oregon, Eugene, Oregon 97403, USA }


\date{\today}

\begin{abstract}
The ability to characterize the complete quantum state of light is essential for both fundamental and applied science. For single photons the quantum state is provided by the mode that it occupies. The spectral temporal mode structure of light has recently emerged as an essential means for quantum information science. Here we experimentally demonstrate a self-referencing technique to completely determine the pulse-mode structure of single photons by means of spectral shearing interferometry. We detail the calibration and resolution of the measurement and discuss challenges and critical requirements for future advances of this method.
%Accurate characterization of the complete quantum state of light is essential for both fundamental and applied science. For single photons the quantum state is provided by the mode that it occupies. The spectral temporal mode structure of light is coming to the fore as an essential means for quantum information science. Here we experimentally demonstrate a self-referencing technique to completely determine the pulse-mode structure of single-photon pulse and discuss challenges for future advances in the techniques and critical requirements for further developments.
\end{abstract}

\pacs{}
%\keywords{}

\maketitle

Quantum photonic technology research is increasingly focusing on ultrashort optical pulsed modes due to their high information content and their compatability with integrated optical platforms. The time-frequency (TF) degree of freedom constitutes an infinite Hilbert space \cite{humphreys:14}, allowing an information content per photon limited only by the encoder and detector resolution. Accessing the information contained in both the spectral amplitude and phase domains of ultrafast pulsed modes of quantum light raises the possibility of surpassing the standard quantum limit in precision measurements such as pulse time-of-flight \cite{lamine:08} and atmospheric characteristics \cite{jian2012real}. Furthermore, the freedom to encode quantum information in multiple spectral-temporal modes, and then to recover that information, extends the usefulness of quantum secure information protocols such as quantum key distribution (QKD) \cite{nunn:13}.  However, for these advantages to be exploited, complete and reliable characterization techniques of quantum light pulses are required. 

%All such applications of quantum optical technologies involve three stages: state preparation, state evolution, and ultimately measurement. Quantum state characterization is crucial for developing all three stages. Firstly, the development of nonclassical light sources (and detectors) requires full characterization of the optical field they interact with. Secondly, for quantum technology applications the action of any operation on a quantum state needs to be well characterized, which will in turn require the full characterization of a probe state. Finally, pure state tomography is a vital part of many quantum metrology tasks.
Quantum optical technologies involve three experimental stages: state preparation, state evolution or active manipulation, and ultimately measurement. The ability to accurately characterize the quantum state of light is crucial for developing all three stages of optical quantum technologies. Firstly, the development of nonclassical light sources requires complete characterization of the optical field output to certify the output light matches the desired quantum state \cite{lounis:05}. Secondly, verifying operations that manipulate the quantum state of light requires full characterization of a complete set of probe states \cite{qin:15}. This forms the basis of optical quantum process tomography. Finally, the development of new quantum optical detectors requires the ability to probe the detector response with a complete set of probe states, a technique known as quantum detector tomography. The set of probe states can be verified by first using a previously calibrated detector.

%For ultrashort pulses, the pulse envelope changes on far shorter timescales than the response times of even the best detectors making direct measurement of the temporal intensity infeasible. Moreover, a full reconstruction of a pulse necessitates the knowledge of its spectral or temporal phase, even if the amplitude is known in both domains\cite{weiner2011Book}.
The pulse envelope of ultrashort optical pulses varies on a time scale far shorter than the response time of the best photodetectors, making direct sampling of the temporal intensity of ultrashort optical pulses infeasible. Moreover, full reconstruction of an optical pulse train necessitates knowledge of its spectral or temporal phase even if the amplitude is known in both the time and frequency domains \cite{weiner2011Book}. Although significant progress has been made in characterizing intense ultrashort light pulses using nonlinear optical interactions with a shorter optical pulse, even enabling resolution of the carrier frequency oscillations \cite{walmsley:09}, techniques for single-photon level characterization are in their infancy. Methods to estimate the optical pulse shape of single-photon sources based on interference with well-known reference pulses have been realized, but such approaches require stable, tunable, mode-matched reference pulses \cite{wasilewski:07,brecht:15,qin:15,polycarpou:12}. Such a tunable reference source is difficult to obtain and requires \emph{a priori} information about the pulse to be characterized to ensure proper overlap with the reference. These approaches also require the reference pulse to be scanned in multiple parameters, thus requiring long measurement times and many on-demand copies of the original state. We therefore seek a self-referencing method for characterizing the ultrafast pulse mode structure of single-photon sources without need for reconfiguration of the apparatus.

Spectral phase interferometry for direct electric field reconstruction (SPIDER) is an established technique for characterising ultrashort pulses in the high field regime. The SPIDER protocol involves interfering an ultrafast pulse with a copy of itself to which a constant translation of the spectrum, or spectral shear, has been applied. Information about the spectral phase of the pulse can be recovered through measurements of the spectral interference pattern between the two pulses \cite{walmsley:09}. In previous work with bright pulses, the spectral shear is typically obtained through second harmonic generation or other processes which are nonlinear in the incident optical field. Other methods of ultrafast pulse characterization such as frequency-resolved optical gating (FROG) also rely on such nonlinear processes \cite{trebino:97}. However, in the quantum regime such nonlinear processes are not practical under general conditions due to the difficulty of achieving high probabilities of nonlinear interactions at the single-photon level in a noise-free, easily reconfigurable setting \cite{brecht:11}.

A promising method of obtaining the spectral shift needed for spectral-shearing interferometry is electro-optic temporal phase modulation \cite{dorrer:03}. This involves passing the light through an electro-optically responsive material such as lithium niobate, in which the refractive index is simultaneously varied such that a temporally varying phase is applied to the pulse. The effect of linear temporal phase modulation is the desired uniform frequency shift. This method has the advantage that it is deterministic and largely independent of the incident pulse shape or amplitude, and is thus applicable at the single-photon level. Recently, the feasibility of such scheme has been shown at the single-photon level \cite{wright:17}.

Previous work has demonstrated the feasibility of electro-optic spectral shearing intereferometry (EOSI) with classical light \cite{dorrer:03}. We demonstrated the first single-photon pulse reconstruction by EOSI \cite{PRL:18}. Here, we provide a detailed description of the methods and thoroughly discuss the scope and challenges of this class of characterization scheme.

\section{Background}\label{background}
\begin{figure}
	\includegraphics[width=.75\linewidth]{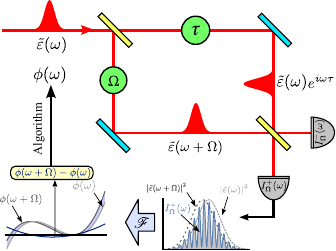}
	\caption{Schematic of EOSI. A pulse is coupled into a Mach-Zehnder interferometer where one arm applies a spectral shear via electro-optic modulation, whilst the other applies a relative temporal delay. Spectrally-resolved detection at the output allows extraction of the spectral phase through Fourier analysis.}
	\label{fig:schematic}
\end{figure}

The quantum state of a single photon is given by the field mode that it occupies \cite{birula:96, sipe:95, smith:07}. Here we focus on the pulse modes of an optical beam, in which the source emits single photons into a well-defined transverse spatial and polarization mode, such as that found in a single-mode optical fiber. To understand how EOSI enables the reconstruction of the pulse mode structure of a single-photon source, we begin by illustrating how the spectral shearing interferometry algorithm extracts the spectral phase of an electromagnetic field mode \cite{kosik:02,kosik:14,dorrer:01,walmsley:09}. We will assume that successive pulses are identical, with the same polarization and spectral mode. We will work in terms of the spectral representation of the analytic field
\begin{equation}
\tilde{\varepsilon}(\omega) = \sqrt{I(\omega)}e^{i\phi(\omega)}=\mathcal{FT}\left[\varepsilon(t)\right],
\end{equation}
where $\mathcal{FT}$ refers to the Fourier transform, $I(\omega)$ is the spectral intensity, $\varepsilon(t)$ and $\tilde{\varepsilon}(\omega)$ are complex-valued analytic functions, $\phi(\omega)$ is the spectral phase and the electric field $E(t)$ is related to $\varepsilon(t)$ by
\begin{equation}
E(t)=\mbox{Re}\{\varepsilon(t)\}.
\end{equation}
Clearly, the electric field $\tilde{\varepsilon}(\omega)$ can be fully reconstructed from measurements of $I(\omega)$ and the spectral phase $\phi(\omega)$. Measurement of $I(\omega)$ can be obtained using a single-photon spectrometer described in section \ref{FBG}. However the spectral phase is more difficult to obtain and measurements must be made indirectly.\\

The EOSI protocol for obtaining the spectral phase works by splitting the pulse with a 50:50 beam splitter, applying a spectral shift to one of the copies and a temporal shift to the other, recombining the two at a second beam splitter, and analysing the spectral interference pattern (see fig.\ref{fig:schematic}). If the spectral shear is given by $\Omega$, the relative time delay by $\tau$ and the intensity of the spectral interference pattern from the two outputs by $I^+_\Omega(\omega)$ and $I^-_\Omega(\omega)$, then 
\begin{eqnarray}
I_\Omega^{\pm}(\omega) &=& \frac{1}{2}|\tilde{\varepsilon}(\omega)e^{i\omega\tau} \pm \tilde{\varepsilon}(\omega+\Omega)|^2 \\ \nonumber &=& \frac{1}{2}(I(\omega)+I(\omega+\Omega)) \label{eqIntensityPattern}\\ \nonumber & & \pm \mbox{Re}\{\tilde{\varepsilon}(\omega)\tilde{\varepsilon}^*(\omega+\Omega)e^{i\omega\tau}\}.
\end{eqnarray}

One can then take the Fourier transform of this intensity pattern:

\begin{eqnarray}
&&\overline{I}_\Omega^{\pm}(T) \equiv \mathcal{FT}\{I_\Omega^{\pm}(\omega)\} \nonumber \\
& = &\frac{1}{2}\overline{I}(T)(1+e^{iT\Omega})\\ \nonumber& \pm & \mathcal{FT}\{|\tilde{\varepsilon}(\omega)\tilde{\varepsilon}(\omega+\Omega)|\exp[i \Delta\phi(\omega,\Omega)]\}*\delta(T-\tau)\\ \nonumber& \pm & \mathcal{FT}\{|\tilde{\varepsilon}(\omega)\tilde{\varepsilon}(\omega+\Omega)|\exp[-i\Delta\phi(\omega,\Omega)]\}*\delta(T+\tau)
\end{eqnarray}
where $*$ indicates a convolution, $\Delta\phi(\omega,\Omega)=\phi(\omega)-\phi(\omega+\Omega)$ and $\overline{I}(T) \equiv \mathcal{FT}\left[I(\omega)\right]$, and so has a width determined by the transform-limited duration of pulses with the spectrum $I(\omega)$.\\

If $\tau$ is chosen to greatly exceed the pulse duration, then the three terms of the above expression can be resolved from one another, and it is possible to computationally filter one of the two side bands, e.g. 
\begin{equation}
|\tilde{\varepsilon}(\omega)\tilde{\varepsilon}(\omega+\Omega)|\mbox{exp}\{i(\omega\tau+\Delta\phi(\omega,\Omega))\}
\label{interm}
\end{equation}

Taking the argument of this yields 
\begin{equation}
\omega\tau+\Delta\phi(\omega,\Omega) \equiv \theta(\omega,\Omega,\tau)
\label{Phi}
\end{equation}
from which the spectral phase can be faithfully reconstructed if $\tau$ is well-calibrated. Up to an irrelevant constant phase, eq.\ref{Phi} can be re-written as 
\begin{equation}
\theta(\omega,\Omega,\tau)=(\omega-\omega_0)\tau+\phi(\omega)-\phi(\omega+\Omega)
\label{deltaphi}
\end{equation}
where $\omega_0$ is some angular frequency, taken henceforth to be the center frequency of the interference term. We call this term the phase gradient, as it is similar, at the first order, to the derivative of the phase with respect to frequency.

The Fourier analysis method used to extract the envelope and the phase from the interference pattern is similar to a Hilbert transform. It is achieved by digitally filtering one sideband in the Fourier domain with a narrow filter and taking either the modulus or the argument of the inverse Fourier transform. This procedure is depicted using experimental data by Fig. \ref{fringes-algo}.

\subsection{Phase reconstruction}\label{sec:algo}
There are several mathematical approaches to extracting the spectral phase $\phi(\omega)$ from the measured phase gradient $\theta(\omega)$. Although in principle any of these algorithms suffices to reconstruct any spectral phase, a different reconstruction method might be appropriate depending on what the spectral phase profile is expected to look like.  
For many applications, only the low-order terms in $\omega$ of the spectral phase $\phi(\omega)$ are of interest and so it is most appropriate to proceed by finding a polynomial expansion of $\phi(\omega)$. 
It is helpful to reformulate eq.\ref{Phi} in terms of a Taylor series expansion of the second argument about $\omega$,
\begin{equation}
\theta(\omega,\Omega,\tau)=(\omega-\omega_0)\tau - \Omega\frac{d\phi}{d\omega}-\frac{\Omega^2}{2}\frac{d^2\phi}{d\omega^2}-\cdots
\label{powerseries}
\end{equation}
Now we express $\phi(\omega)$ as a power series about $\omega_0$ and defining $\delta\omega \equiv (\omega-\omega_0)$,
\begin{equation}
\phi(\omega_0+\delta\omega) = \phi_0+\phi_1\delta\omega+\phi_2\delta\omega^2+\phi_3\delta\omega^3\cdots
\end{equation}
where $\phi_n$ is the $n$-th term in the expansion for the spectral phase. Substituting into eq.\ref{powerseries} we obtain
\begin{eqnarray}
\nonumber \theta(\omega,\Omega,\tau)=\delta\omega~\tau&-&\Omega(\phi_1+2\delta\omega\phi_2+3\delta\omega^2\phi_3\cdots) \label{linear}\\&-&\Omega^2(\phi_2 +3\phi_3\delta\omega +\cdots)\\&-&\Omega^3(\phi_3+\cdots \nonumber
\end{eqnarray}
At this point we remark that if $\Omega$ is small compared to the bandwidth of the pulse, then for most of the frequency range where $\theta(\omega,\Omega,\tau)$ is large compared to its error (i.e., the spectral overlap region of the pulses) eq.\ref{powerseries} reduces to
\begin{equation}
\theta(\omega,\Omega,\tau)=\tau\delta\omega - \Omega\frac{d\phi}{d\omega}
\end{equation}
and hence $\phi(\delta\omega)$can be directly extracted as
\begin{equation}
\phi(\delta\omega)=\frac{1}{\Omega}\int [\tau\delta\omega-\theta(\omega,\Omega,\tau)]d\omega
\label{integral}
\end{equation}
with the integration performed over the region where $\theta(\omega,\Omega,\tau)$ can be reliably measured. However, eq.\ref{linear} shows that in this limit
\begin{equation}
\theta(\omega,\Omega,\tau)=-\Omega\phi_1 +\delta\omega(\tau-2\Omega\phi_2)-6\Omega\delta\omega^2\phi_3+\cdots
\label{chirpinduceddelay}
\end{equation}
and hence that any error in the proper calibration of $\tau$ manifests after the integration over $\omega$ as an error in the extracted value of $\phi_2$, the second-order spectral phase or ``chirp". It is therefore vital that $\tau$ is accurately characterized, with error much less than $2\Omega\phi_2$. 

\begin{figure*}[t!]
	\centering
	\includegraphics[width=.6\linewidth]{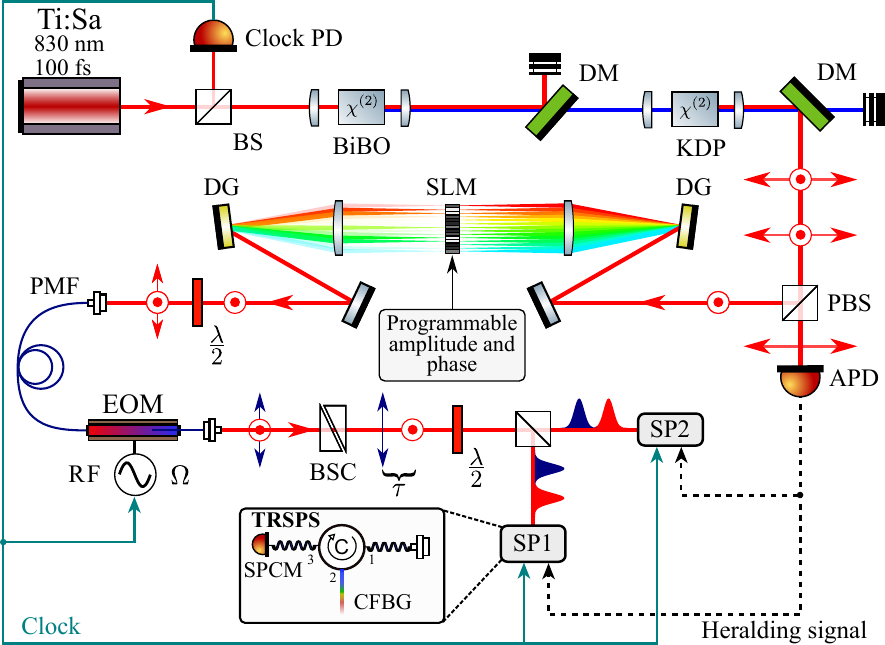}
	\caption{Experimental setup for single-photon generation, shaping and EOSI detection (see text for detailed description). Ti:Sa: Titanium-Sapphire femtosecond oscillator; BS: beamsplitter; PD: photodiode; DM: dichroic mirror; PBS: polarizing beam-splitter; APD: single-mode-fiber-pigtailed avalanche photodiode; DG: diffraction grating; SLM: spatial light modulator; PMF: polarization-maintaining fiber; EOM: electro-optic phase modulator; RF: radio-frequency signal; BSC: Babinet-Soleil compensator; $\lambda/2$: half-waveplate; SP: spectrometer; CFBG: chirped fiber Bragg grating; SPCM: single photon counting module; TRSPS: time-resolved single-photon spectrometer.}
	\label{fig:setup}
\end{figure*}
Within the regime of the shear $\Omega$ being much less than the bandwidth of the pulse, the errors in the phase reconstruction will be smaller with increasing shear. If the shear were comparable to the pulse bandwidth, the visibility of interference fringes will be lost due to the reduced overlap of the sheared and unsheared pulses, which would negatively affect phase recontruction. The spectral shear applied by the set-up can be tuned up to a maximum of approximately 0.3 nm, for a pulse bandwidth of several nanometers, and so the interferometer operates comfortably within the regime where the overlap is close to unity. However the shear is large enough relative to the bandwidth that the successive terms in eq.\ref{linear} contribute significantly across much of the range of the integral in eq.\ref{integral} and may not be ignored. Therefore, to extract the phase terms, we note that eq.\ref{linear} is linear in the spectral phase terms $\phi_i$ and hence may be expressed
\begin{equation}
\delta\omega~\tau-\theta(\delta\omega,\Omega)=A\vec{\phi}
\label{A}
\end{equation}
where $\vec{\phi}$ is the vector $(\phi_1,\phi_2,\phi_3 \cdots)^T$ and $A$ is a matrix defined with its columns as
\begin{equation}
A\equiv(\Omega,(\delta\omega+\Omega)^2-\delta\omega^2,(\delta\omega+\Omega)^3-\delta\omega^3,\cdots)
\end{equation}
In our analysis we chose $A$ to include terms up to $(\Omega/\delta\omega)^4 \approx 10^{-4}$ to keep calculation errors small.  Eq.\ref{A} can be efficiently and accurately solved numerically using a matrix division routine to finally obtain the spectral phase.

For arbitrary spectral phase (provided the pulse's temporal support is limited to the region $[-\pi/\Omega, $ $\pi/\Omega]$), the phase can also be recovered through a concatenation approach \cite{dorrer:01}. This method is more suitable for spectral phase profiles that are non-analytic in $\omega$ about the center frequency $\omega_0$ or contain high-order polynomial contributions in $\omega$ that would not be reconstructed by the above approach. This approach begins by setting the spectral phase at some value $\phi(\omega_0)$ to be some constant arbitrary value, typically zero, and then iteratively reconstructing the phase at every point $\omega+n\Omega$ for integer $n$, such that
\begin{equation}
\phi(\omega_0+[n+1]\Omega)=\phi(\omega_0+n\Omega)+\theta(\omega_0+n\Omega)-\Omega\tau
\label{eq:concat}
\end{equation}

According to the Shannon theorem, this concatenation approach suffices to completely characterise the pulse if it has no temporal support outside the range $[-\pi/\Omega, \pi/\Omega]$ \cite{walmsley:09}. If this condition is not met, however, a disadvantage of this method is that it only provides the phase relationships between points separated by an integer multiple of the shear, which is often a coarser sampling than that given by the resolution of the spectrometers. Therefore, whilst this process gives the exact phase relationships of discrete sets of points separated by integer multiples of the shear, the phase differences between points offset by amounts other than an integer multiple of the shear is not directly determined. One approach to solving this is to use each series to determine the profile of a mode in the time domain, then average these pulses. The resulting pulse profile contains information from all the contributing frequency measurements, and can be Fourier transformed back into the spectral domain to provide the complete set of phase relationships \cite{dorrer:01}.

When processing the EOSI interferogram, care must be taken since most spectrometers will sample the intensity profile in the wavelength, rather than frequency, domain. For narrowband pulses it is acceptable to assume that the bandwidth of the pulse is sufficiently smaller than the wavelength that a linear relationship is sufficient to translate the spectrum from the wavelength domain to the frequency domain. However this approximation is not valid in our demonstration and thoroughly corrupts the extracted spectral phase if implemented \cite{dorrer:10}. In our implementation we corrected for the problem by using Fourier transform interpolation, which functions well even up to fringe spacing at the Nyquist limit.

\section{Source of shaped single photons}
The full experimental setup is depicted by Fig. \ref{fig:setup}. The laser source is a commercial femtosecond oscillator (SpectraPhysics Tsunami) delivering pulses of 100 fs full width at half maximum (FWHM) at a repetition rate of 80 MHz, corresponding to a spectrum centered at 830 nm and a bandwidth of 10 nm FWHM. An internal fast photodiode is used to generate from the repetition rate the clock signal that is used throughout the experiment. Second harmonic generation is achieved in a 1mm-long BiBO crystal, generating a 3nm FWHM spectrum at 415nm. Heralded single photons are generated by collinear, type-II spontaneous parametric down conversion (SPDC) in an 8mm-long potassium dihydrogen phosphate (KDP) crystal \cite{mosley:08}. The orthogonally-polarized signal and idler fields with, respectively, 3 nm and 12 nm bandwidths are separated at a polarizing beam splitter (PBS), with a photodetection event in the idler mode heralding the creation of a photon in the signal mode. The phase-matching of the source was chosen such that the complex joint amplitude of the two-photon state produced by the source would be separable, such that detection of a heralding photon in the idler arm projects the signal into a spectrally-pure single-photon Fock state \cite{mosley:08b}. Double pairs of photons, which would cause problems at the output of the interferometer, are shown to be negligible compared to single photon events ($g^{(2)} = 0.06 \pm 0.04$).

To test the EOSI with a range of spectral phase profiles, the signal photon is then directed to a fiber-coupled pulse shaper capable of performing arbitrary spectral phase operations \cite{weiner:00}. The shaper consists of a standard 4-f line built with a 2000 lines/mm diffraction grating (Spectrogon) and 200 mm focal length cylindrical lens (Thorlabs) with a 2D phase mask (Hamamatsu SLM, 1272 x 1024 pixel mask) and is capable of achieving arbitrary spectral phase shaping with a resolution of 0.04 nm/pixel and near-uniform spectral intensity transmission of $\approx$40\%. Losses are equally distributed between the diffraction grating efficiency and the insertion losses in fiber coupling at the output of the device. 

The shaper is calibrated by leaving the phase profile of the SLM uniform except for a single pixel with a $\pi$ phase shift scanned across the mask. This creates a spatial discontinuity in the phase over an aperture comparable in size to the wavelength, which causes diffraction at the wavelength corresponding to the pixel position. The pixel position of the SLM can therefore be mapped to the wavelength by sending in bright classical pulses through the SLM and then into a conventional spectrometer (Shamrock 303i monochromator, 1200 mm$^{-1}$ grating, Andor) and then monitoring the position of the dip in the spectral intensity caused by diffraction around the phase-shifted aperture. 

Since the shaper itself introduces spectral phase, it needed to be calibrated. The exact phase introduced by the shaper in its unmodulated state was determined by standard spectral interferometry. A second, free-space beam path was built around the pulse shaper to form a Mach-Zehnder interferometer with the shaper in one arm, into which bright classical pulses were sent. The relative spectral phase in the two arms introduced by the pulse shaper was then extracted using Fourier algorithms\cite{weiner2011Book}. This was eventually used to create a default mask for the SLM, designed to compensate for the nonuniform spectral phase introduced by the device itself and to therefore allow complete control and manipulation of the spectral phase. 

The SLM is polarization-sensitive, so it is important that only light linearly polarized in the correct orientation is sent into the pulse shaper. However, due to slight misalignment and imperfection in the grating, a small amount of the other polarization appears in the line. A polarizing beamsplitter was added to remove the unwanted polarization and to ensure that no interference fringes due to polarization, which could affect the phase reconstruction, are observed at the output of the pulse shaper. Note that the device is also capable of adjusting the spectral amplitude \cite{Weiner2011} which is used to reduce the bandwidth by the idler photon from 12 to 8 nm in order to fit within the acceptance window of the fiber Bragg grating used in our single-photon spectrometer (see Sec.\ref{FBG}).

\section{Obtaining spectral shear}

To obtain the spectral shear we used an EOSpace LiNbO$_3$-waveguide electro-optic phase modulator (EOM). The modulator is driven by a 10 GHz sinusoidal voltage emitted by a parametric dielectric resonant oscillator (PDRO) and amplifier (Aspen Electronics). The PDRO upconverts by a factor of 125 the 80 MHz signal from the laser cavity photodiode, ensuring that the 10 GHz output signal is phase-locked to the optical pulse train. The modulator is designed such that an applied radio-frequency (RF) signal propagates through the device with phase velocity equal to the group velocity of an optical pulse centered at 830 nm wavelength. A pulse travelling through the device therefore acquires a time-dependent phase $\phi(t)$, depending on the temporally locked RF field amplitude with which it co-propagates. This results in a temporal variation in the refractive index of the waveguide viewed from the reference frame of the optical pulse. The voltage difference necessary to achieve a relative phase shift of $\pi$ between orthogonal polarization propagating through the EOM is referred to as the half-wave voltage $V_\pi$ and depends on the length of the modulator and the electro-optic coupling at a given frequency. \\
The time-varying refractive index of the waveguide can then be used for deterministic manipulation of the frequency spectrum of the pulse, such as spectral shear \cite{wright:17} and bandwidth manipulation \cite{karpinski:17}. If the centre of the pulse co-propagates with the steepest part of the sinusoidal RF waveform, then the optical field at the output of the modulator may be written
\begin{equation}
\varepsilon(t)=\varepsilon_0(t) \mbox{exp}\left(- i\pi \frac{V_{\mbox{max}}}{V_\pi}\sin(2\pi ft)\right),
\end{equation}
where $V_{\mbox{max}}$ is the maximum voltage of the applied RF field, $f$ is the frequency of the RF signal and $\varepsilon_0(t) \approx |\varepsilon_0(t)|e^{i\omega_0 t}$ is the temporal profile of the incident pulse where $\omega_0$ is the central frequency of the wavepacket. If the pulse is short compared to the period of the modulation (in our case, 100 ps), then we can neglect higher-order terms in the expanded temporal phase, resulting in
\begin{equation}
	\varepsilon(t) = |\varepsilon_0(t)|\mbox{exp}\left(i\left(\omega_0 - \frac{2\pi^2 fV_{\mbox{max}}}{V_\pi} \right)t\right)\label{linearregion}
\end{equation}
The result is therefore an effective frequency shift of $\Omega = 2\pi^2 fV_{\mbox{max}}/V_\pi$. Since the temporal support of the pulse must be within the region where the RF signal is linear, there is a limit on the maximum temporal duration of pulses to which a uniform shift can be applied without modifying the spectral envelope. 

In practice, the phase-lock between the RF signal and the optical pulse train will have some noise, resulting in each pulse experiencing timing jitter with respect to the RF signal. However, the effect of this is simply to reduce the visibility of the fringes in the interferogram, and does not systematically affect the calculated spectral phase \cite{dorrer:08}.

The electro-optic response of the medium is much greater in the crystal axis aligned with the applied electric field. For light polarized along the propagation direction, the coefficient of the electro-optic tensor has a magnitude of 33 pm/V, while their values are $\approx 3$ pm/V for other polarization. Hence, light propagating in the orthogonal polarization is less affected since the modulation of the index of refraction is directly proportional to the electro-optic coefficient. By propagating the unshifted pulse through the modulator in an orthogonal polarization, it is possible to achieve a common spatial path for the two pulses \cite{dorrer:10}. This common path approach has the advantage of making the set-up more resistant to vibrational and thermal fluctuations in path length and so improving interferences visibility, as well as minimizing the relative spectral phase between the two arms of the interferometer. Overall in our set-up we were able to achieve a relative spectral shear of approximately 0.58 nm at 830 nm central wavelength.

Generally, when the RF source is turned on, it will lock to an arbitrary phase of the pulse train, meaning the pulse co-propagates at a random point of the RF waveform. Furthermore, when the source is active, the phase lock is dependent on thermal fluctuations which could result in a slow drift of the lock point. It is therefore necessary to stabilize the temperature environment. The amplitude and phase of the RF signal wave can be adjusted in real time through a variable time delay and attenuator \cite{wright:17}. For bright classical light, it suffices to send the beam (of the modulated polarization) into the EOM and spectrally resolve the output, and then select the setting of the RF phase control such that the center wavelength of the pulse is extremized. However, the single photon signal produces far fewer counts, making reliable alignment in real time vastly more challenging. To make the process easier, a bright coherent beam line is coupled into the same beam path as the downconverted light during the photon-source alignment. This bright classical beam line was constructed such that the delay in the pulse train relative to the single-photon pulses was an exact integer multiple of the RF waveform period (100 ps). By setting the delay to precisely 200 ps, it is known that the bright coherent pulses are aligned with the same point of the RF waveform as the single photon pulses, and therefore will receive the same spectral shear. In this way it is possible to ensure the single photons are spectrally shifted by inspecting the spectrum of bright classical light.

\section{Calibrating the interferometer}

Our implementation of the EOSI uses a polarization Mach-Zehnder configuration. This has several advantages over a spatial mode Mach-Zehnder or other interferometer designs: lower losses, spatial-mode matching, common-path stability, and a reduced need to account for internal spectral phase differences within the interferometer. One minor disadvantage is that the unmodulated pulse also passes through the modulator, making it susceptible to residual electro-optic effects. This could affect the relative shear $\Omega$ or otherwise distort the reference pulse and thus corrupt the extracted spectral phase, although this proved negiligible in our demonstration. Several factors reduce the visibility of the interferogram: thermal fluctuations in the interferometer, the timing jitter of the RF signal with respect to the optical pulse train,  the mixedness of the single-photon state arising from optical fluctuations in the pump and pulse shaper, and mixedness arising from imperfect separability of the two-photon state and the tracing over the state of the herald photon. Besides optical effects, the limited resolution caused by the point-spread function of the spectrometers also creates the appearance of reduced visibility.

The value of $\tau$ must be chosen such that the spectral fringe spacing is well below the sampling limit of the spectrometer (corresponding to $\tau \approx 27$ ps) to provide an interferogram with good visibility, but large enough such that the interference term is clearly separable from the DC signal and that spectral fringes appear with sufficient density to extract the phase to high order. In the experiments, a time delay of $\tau \approx 5$ ps was chosen. 

After passing through the pulse shaper, the signal photon is prepared by a half wave plate, acting as the input of the polarization Mach-Zehnder interferometer (PMZI), at 45$^\circ$ polarization to the optical axis of the PM-fiber coupled electro-optic modulator into which it is directed. The single photon is thus in a superposition of polarization states, one of which experiences a constant spectral translation (the EOM applies only a marginal spectral shear on the ordinary ray). The pulses were then directed into a free-space path containing birefringent calcite wedges (a Babinet-Soleil compensator, or BSC), which introduce a variable relative delay between the two polarizations. In practice the PM fiber, which is highly birefringent, and the modulator itself introduce more than enough delay to resolve the interference terms and the role of the calcite wedges is to easily set the spectral interference fringe spacing to below the sampling limit of the spectrometer. The two pulses are then rotated in polarization by a final half wave plate oriented at 22.5$^\circ$ to their polarizations and interfered at a polarizing beam splitter (closing the PMZI), with the resulting spectral interference patterns resolved on time-resolved single photon spectrometers (see Sec.\ref{FBG}).\\

\begin{figure}
	\includegraphics[width=.85\linewidth]{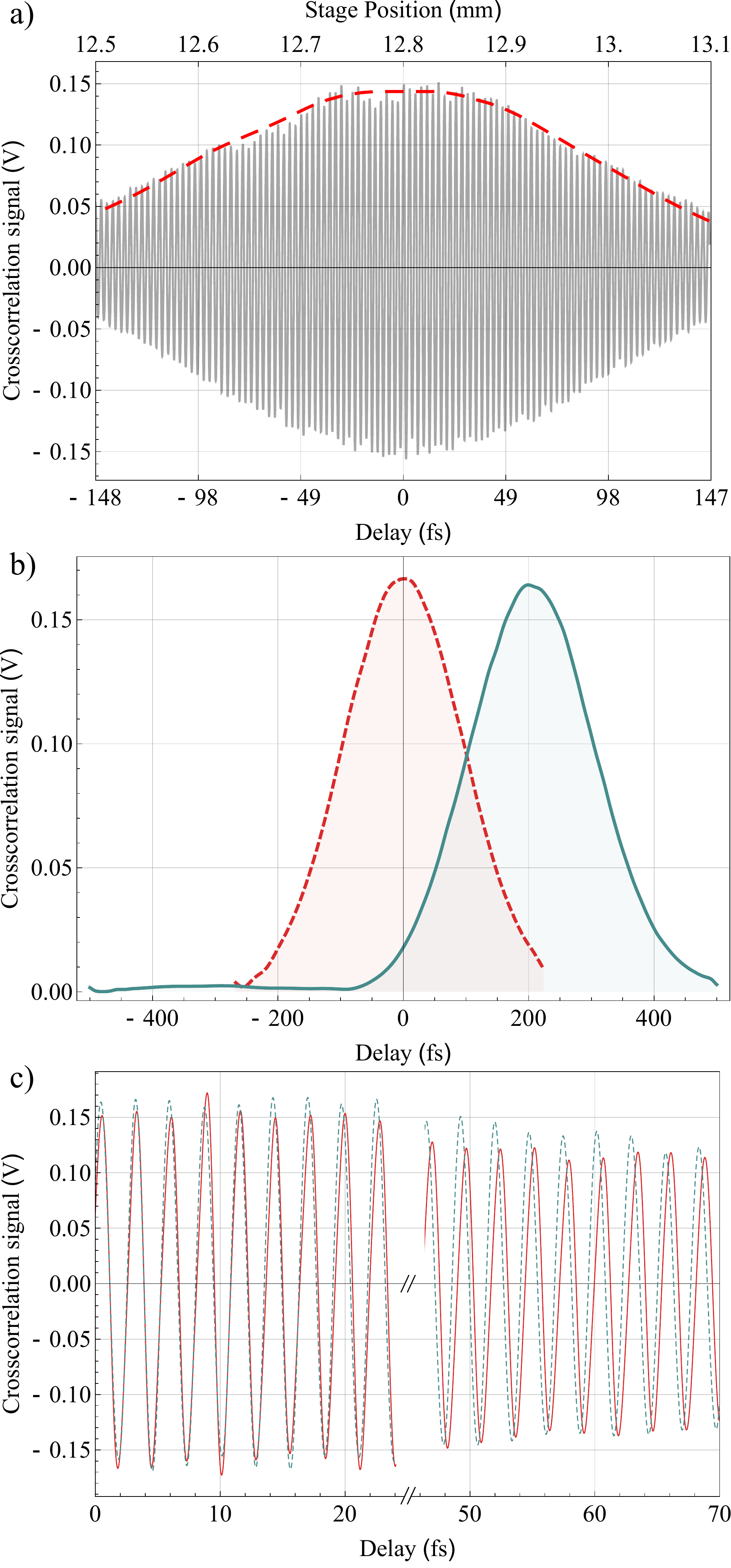}
	\caption{a) Temporal cross-correlation pattern of unsheared classical wavepackets with the calibrated time axis; dashed: envelope of the cross-correlations obtained by Hilbert transform.
		b) Comparison of the envelope of the cross-correlation pattern for an unsheared (dashed) and sheared (plain) classical wavepacket, showing a delay introduced by the group velocity change of approximately 200 fs. c) Superposition of carrier-wave fringes showing difference in frequency caused by the spectral shear $\Omega$.}
	\label{fig:calibration}
\end{figure}

To reconstruct the spectral phase, it is necessary to have precise knowledge of certain parameters of the interferometer. In particular the birefringent time delay $\tau$, which must be known to much higher accuracy than the chirp-induced effective delay $\Omega\phi_2$ for accurate reconstruction, and the spectral shear $\Omega$. The value of $\Omega$ can be chosen and measured by sending classical light of the modulated and unmodulated polarizations into the interferometer and resolving the spectrum. However, the value of $\tau$ is more challenging to determine and much more important to the accuracy of the reconstruction. The error of the value of $\tau$ that would lead to errors in the reconstructed phase comparable to the typical values one would wish to measure is on the order of several hundred femtoseconds, out of an absolute value of approximately 5 ps. It is therefore not possible to perform direct measurements of $\tau$ to satisfactory accuracy with a conventional photodiode. Attempts to calculate the value of $\tau$ theoretically are challenging due to the several processes within the modulator that contribute to the delay. The main contribution to $\tau$ is from the birefringence of the LiNbO$_3$ in the modulator, which is strongly temperature dependent. Furthermore, LiNbO$_3$ is a highly dispersive material, and so the difference in group velocity of the sheared and unsheared pulses and, again, the temperature dependence thereof, would need to be taken into account. One would also need to simulate how the group velocity of the sheared pulse varied as its spectrum was translated during its propagation through the modulator. We therefore used an interferometric means of directly measuring $\tau$.

Calibration of the calcite wedges is done by injecting bright classical pulses derived from the laser into the interferometer with the EOM switched off and measuring with a silicon photodiode the intensity at the output whilst scanning the position of the calcite wedges. This method only requires knowledge of the carrier frequency of the incident light to map the wedge position to delay. Firstly it is necessary to create a mapping of wedge position onto relative time delay introduced by the birefringence of the calcite. The interference signal at one output of the interferometer is given by:
\begin{equation}
	I_\textrm{out} = \frac{1}{2} \int \textrm{d}\omega ~ I_0(\omega) \cos{\left(\phi_\textrm{rel}(\omega)\right)}
\end{equation}
where $I_0$ is the spectral intensity of the incident field and $\phi_\textrm{rel}$ is the relative spectral phase within the PMZI. With the EOM off, the relative phase $\phi_\textrm{rel}(\omega)$ only contains a linear term that corresponds to a global delay $\tau$ between the polarizations, and so it is possible to approximate the signal as:
\begin{equation}
	I_\textrm{out}(\tau) = I_\textrm{cc}(\tau) ~ \cos{(\omega_0 \tau)}
	\label{xcorr}
\end{equation}
where $I_\textrm{cc}$ is the envelope of the cross-correlation and $\omega_0$ is the carrier frequency of the pulse. It contains all the delay from propagation through the EOM and the polarization-maintaining fiber, and is also directly proportional to the length of calcite through which light travels. The latter allows to control the delay between the two polarizations. Hence, the first moment of the envelope gives the point of zero delay, while the fringe spacing gives the temporal delay as a function of wedge position, as shown by Fig. \ref{fig:calibration}a). To maximize overlap between the pulses in the two polarizations, this time-axis calibration would ideally be done with the EOM switched off. However, due to the aforementioned thermal effects on the refractive indices of the modulator, the axes can only be calibrated across the range of interest with the EOM being driven. Therefore, the phase of the RF relative to the optical pulse train was chosen such that the pulse sits at the extremum of the refractive index modulation. This leaves the center wavelength of the pulse unchanged, allowing extraction of the time delay by the method described above. Note that any change in fringe spacing due to mismatch of higher order spectral phase, such as dispersion, will only appear in the wings of the interferogram, and can be easily accounted for with Fourier analysis. This method then allows to unambiguously calibrate the temporal delay caused by the calcite wedge. 

With the axis calibration in hand it is then possible to extract the time delay $\tau$ of the interferometer. Since the pulse reconstruction experiment involves spectrally shearing one arm within the interferometer, this results in a change in the time delay $\tau$ between the two polarizations due to group delay dispersion. In this case, the cross-correlation signal is given by
\begin{equation}
I_{cc}(\tau)=I_\textrm{cc}(\tau-\tau_0)\mbox{cos}\left[(\omega_0+\Omega/2)(\tau-\tau_0)\right]
\end{equation}
where $\tau_0$ is the contribution to the delay caused by group delay dispersion of the sheared pulse as derived in eq.\ref{chirpinduceddelay}. The interferogram measured when the EOM is shearing is then delayed by a constant value $\tau_0$, and the difference in fringe spacing also allows measurement of the spectral shear. As shown in figure \ref{fig:calibration}b), the zero-delay point is shifted by about 200 fs when maximum shear is applied on the EOM. The sign of this delay depends on the sign of the shear (positive for lower wavelength), which is consistent with the index progression ($n$ decreases when $\lambda$ increases). By defining the zero delay point as the new reference, the stage position is then fully calibrated and the time delay $\tau$ between the two arms of the interferometer is fully characterized.  The advantage of this calibration method is that it does not require any ultrafast detection: the sampling rate of the acquisition simply has to be sufficient to resolve temporal fringes created by the movement of the stage, which is typically have a period on the order of a few kHz. It does however require phase stability, but this is obtained directly from the common-path polarization interferometer.
Finally, as depicted by fig.\ref{fig:calibration}c), when the two interferograms (shear vs no shear) are superimposed, the difference in optical frequency due to the difference in wavelength can be clearly seen. Performing the scan multiple times and averaging the phase extracted from the Fourier analysis conclusively allows the extraction of the shear $\Omega$. Though directly measuring the shear using a spectrometer is more robust, in the absence of a high-resolution conventional spectrometer the cross-correlation method allows a direct measurement at the output of the interferometer, taking into account every optical element. 

\begin{figure*}[ht!]
	\includegraphics[width=.9\linewidth]{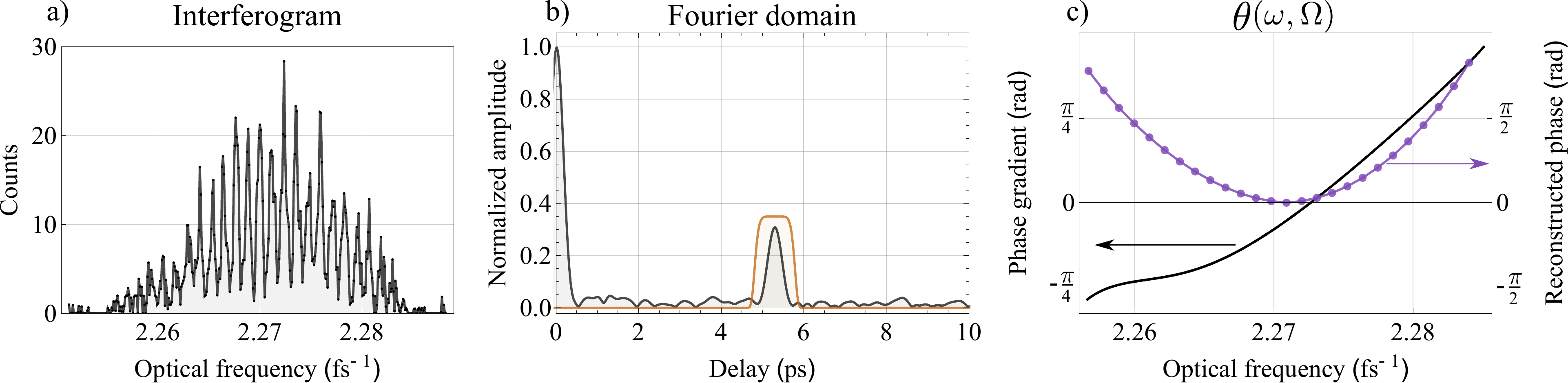}
	\caption{Phase extraction process for an unshaped heralded single photon. a) output of the interferometer, showing a 20 seconds acquisition. b) Fourier transform of the interferogram, showing a peak centered around the delay $\tau$ of 5ps. Overlay: filter used to isolate the sideband. c) argument of the inverse Fourier transform of the filtered sideband (black) and reconstructed phase (blue). The delay $\tau$ has been removed, showing a linear slope due to the quadratic phase accumulated through fiber propagation.}
	\label{fringes-algo}
\end{figure*}

\section{Spectrally-resolved detection}
\label{FBG}
Since spectral-shearing interferometry requires spectrally-resolved measurement, it is necessary to simultaneously gather enough single photon counts in multiple spectral channels before contrast is lost in the interferometer. Conventional spectrometers often work by coupling spectral modes into some other basis, such as spatial modes, by means of a prism or diffraction grating, and monitoring each of those modes with a separate detector. However, spectrometers capable of concurrently monitoring multiple spectral modes with single-photon level intensity resolution have been difficult to realize due to the size and expense of arrays of multiple single-photon counting detectors. One solution for short pulsed modes with a well-characterized repetition rate is dispersive Fourier-transform spectroscopy, whereby the pulse undergoes frequency-to-time mapping in a dispersive medium such that the temporal envelope of the output pulse corresponds to the spectral envelope of the original pulse \cite{avenhaus:10,muriel1999,azana2000}. Temporally-resolved detection with just one single photon counting module (SPCM) then provides spectrally-resolved detection. The spectrometers implemented here use extremely high dispersion fiber Bragg gratings to map the frequency of a temporally-short optical pulse onto the temporal envelope of the output pulse. We have recently shown that by using state-of-the-art chirped fiber Bragg grating (CFBG) as the dispersive element and low-timing jitter single-photon detectors, high resolution spectrometers can be built \cite{davis:17}.

The optical component of the TRSPS, consisting of a chirped fiber Bragg grating (Teraxion) spliced onto port 2 of an optical circulator (see inset of fig.\ref{fig:setup}), has a transmission efficiency of approximately 10\%. The CFBG introduces a frequency dependent delay of approximately 950 ps/nm on the photon, which is then detected by a fast SPCM (PDM, Micro Photon Devices) with approximately $\Delta t$ = 30 ps timing resolution. The SPCM signal is processed by a time-to-digital converter (TDC, Pico- Quant, HydraHarp 400) triggered by a fast photodiode signal that samples the laser pulse train acting as the experiment clock. The TDC has timing resolution down to 12 ps, below the detector timing resolution. The trigger photodiode (Alphalas, UPD-15- IR2-FC) has approximately 15 ps rise time and timing variation less than 1 ps. Since the spectral resolution is ultimately limited by the timing resolution of the SPCM, the device is able to detect single photons with approximately 55 pm spectral resolution at 830 nm. The Bragg grating is written to reflect light across a wavelength range of 825--835 nm and therefore for broadband pulses also acts as a bandpass filter.

\section{Results}
	\subsection{Data acquisition}
The procedure for a typical experimental single-photon state characterization is depicted by Fig.\ref{fringes-algo}. An acquisition consists of accumulating a spectrogram at the output of the TRSPS over approximately 20 seconds (see Fig. \ref{fringes-algo}a), which was found to be a sufficiently long acquisition time for precise phase extraction given the final coincidence count rates. The visibility of the interference fringes is typically greater than 70$\%$ and is limited by phase instabilities within the interferometer and the resolution of the spectrometer. Performing the Fourier analysis detailed in Sec.\ref{background} then allows extraction of the phase gradient $\theta$ shown in Fig. \ref{fringes-algo}c from the sideband in the Fourier domain (Fig. \ref{fringes-algo}b). The spectral phase information is finally obtained by following the algorithm from Sec.\ref{sec:algo}. Note that the spectral range over which the phase is reconstructed must be restricted to the region of the spectrum in which fringes are observed, since the extracted phase outside of this region is not recoverable (see the edge of Fig. \ref{fringes-algo}c)

To build statistics, this procedure is repeated for multiple acquisitions, each consisting of 20 seconds of data, and the extracted phases are averaged. However, the individual interferograms themselves may not always be summed because long-term instabilities, such as slow temperature drifts, result in a slow shift of the global phase of the interference fringes and so reduce the contrast of the resultant interferomgram. Interferograms that do not reach a threshold contrast of $50\%$ are discarded and do not contribute to the final reconstruction.

\subsection{Reconstruction of phase that is polynomial in frequency}

The experiment was conducted with various spectral phase profiles imprinted on the pulse shaper in order to test the fidelity of the reconstruction. In the absence of deliberately applied spectral phase, the device recovers a spectral phase with a quadratic coefficient of $87 000 \pm 1000$ fs$^2$ and a cubic of $(5\pm 1) \cdot 10^5$ fs$^3$. This value is found to be in good agreement with the theoretically predicted spectral phase the single photons in our experiment receive from their generation and propagation, taking into account the approximately 2.5 meters of polarization-maintaining fiber, the KDP crystal, and the relative phase within the interferometer (calcite wedges and EOM waveguide). This is referred to as the ``uncompensated phase".

\begin{figure}[ht!]
	\includegraphics[width=1\linewidth]{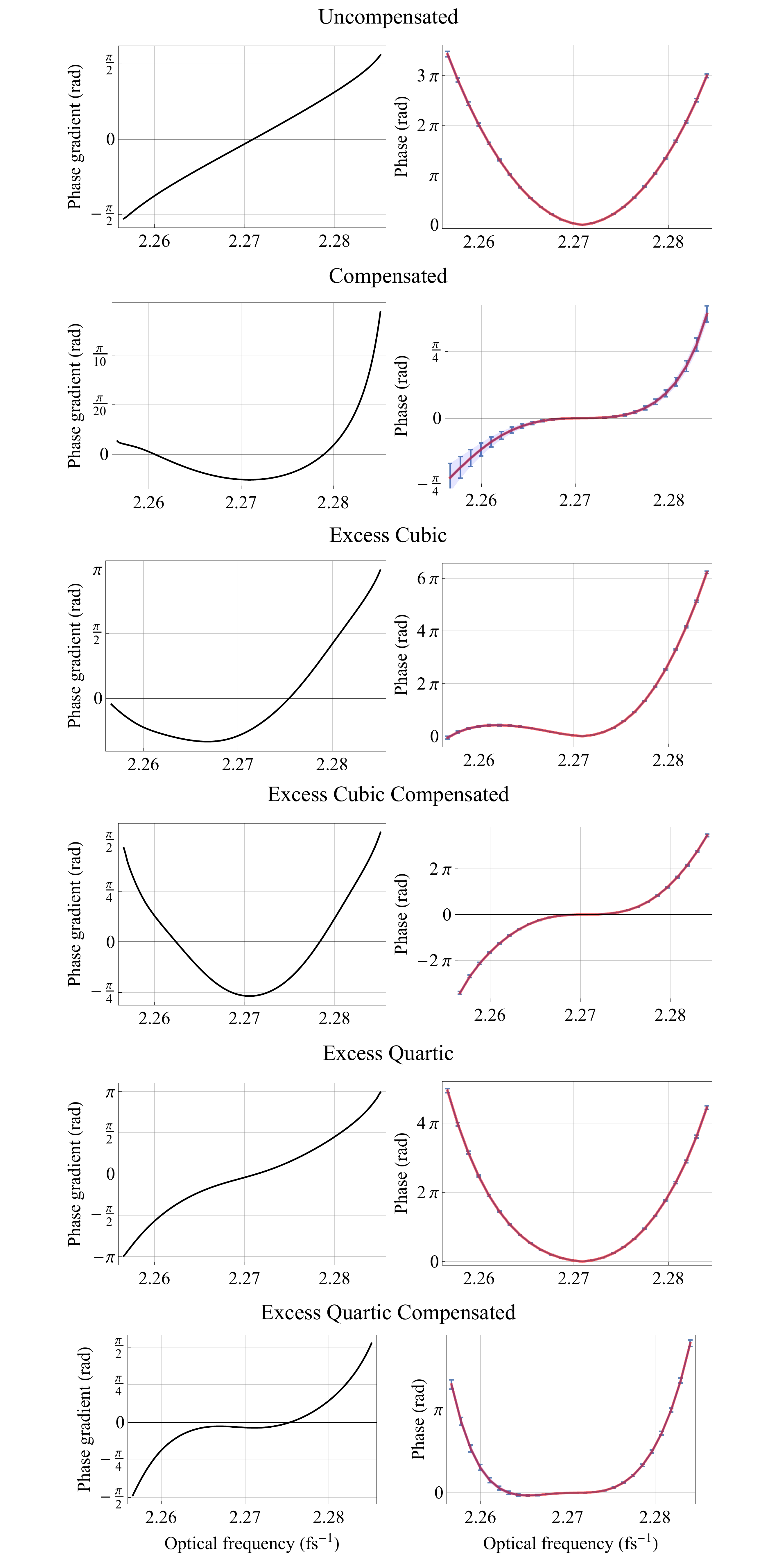}
	\caption{Left column: average of the extracted phase gradients corrected for delay; right column: reconstructed phases from the left column using the concatenation algorithm for a frequency shear $\W<0$. The error bars represent the standard deviation. The error is zero at the center frequency since it is constrained by the algorithm. The left column is approximately the derivative of the right one.}
	\label{poly-phases}
\end{figure}

This quadratic spectral phase is then removed by applying a compensating phase mask on the pulse shaper. The cubic component is not compensated for. Similarly, additional phases are tested, such as excess cubic or quartic phase, with and without the compensation for the second-order phase. The phases that are extracted from the interference pattern are all shown on Fig. \ref{poly-phases} for the different configurations of the pulse shaper, as well as the reconstructed phases. In the first case, without any applied phase on the shaper, we can see that the extracted phase gradient, $\theta(\omega,\Omega)$, is mostly linear and negative and the reconstructed phase itself is quadratic and positive. The difference in sign comes from the sign of the spectral shear, which was negative (\emph{i.e.} positive in wavelength, $\delta \lambda = 0.58$ nm), and extracted phase can be approximated to the first order as the gradient of the real phase, hence a quadratic phase has a linear gradient. This trend can be verified in all the different phases that are imprinted on the pulse shaper. One can also verify that in the compensated case, the remaining phase is quite small in comparison to the large quadratic phase. A cubic component can be extracted, the value of which is found to be equal to that recovered in the uncompensated case, implying the expected additivity of the applied phase.

Polynomial phase profiles up to quartic spectral phase were imprinted on the single photon by the pulse shaper to test the limit of the reconstruction, in every case with or without compensation of the quadratic part. In each case, it can be seen that the reconstruction is consistent with the phase applied on the shaper. The uncompensated cases always show a quadratic part around the same value ($\sim 90 000$ fs$^2$), which falls to zero in the compensated case, while the higher order coefficients remain. The retrieved coefficient for cubic phase is $(2.5 \pm 0.3) \cdot 10^7$ fs$^3$ and $(2.0\pm 1.0) \cdot 10^9$ fs$^4$ for quartic phase.

\begin{figure}[t!]
	\includegraphics[width=.9\linewidth]{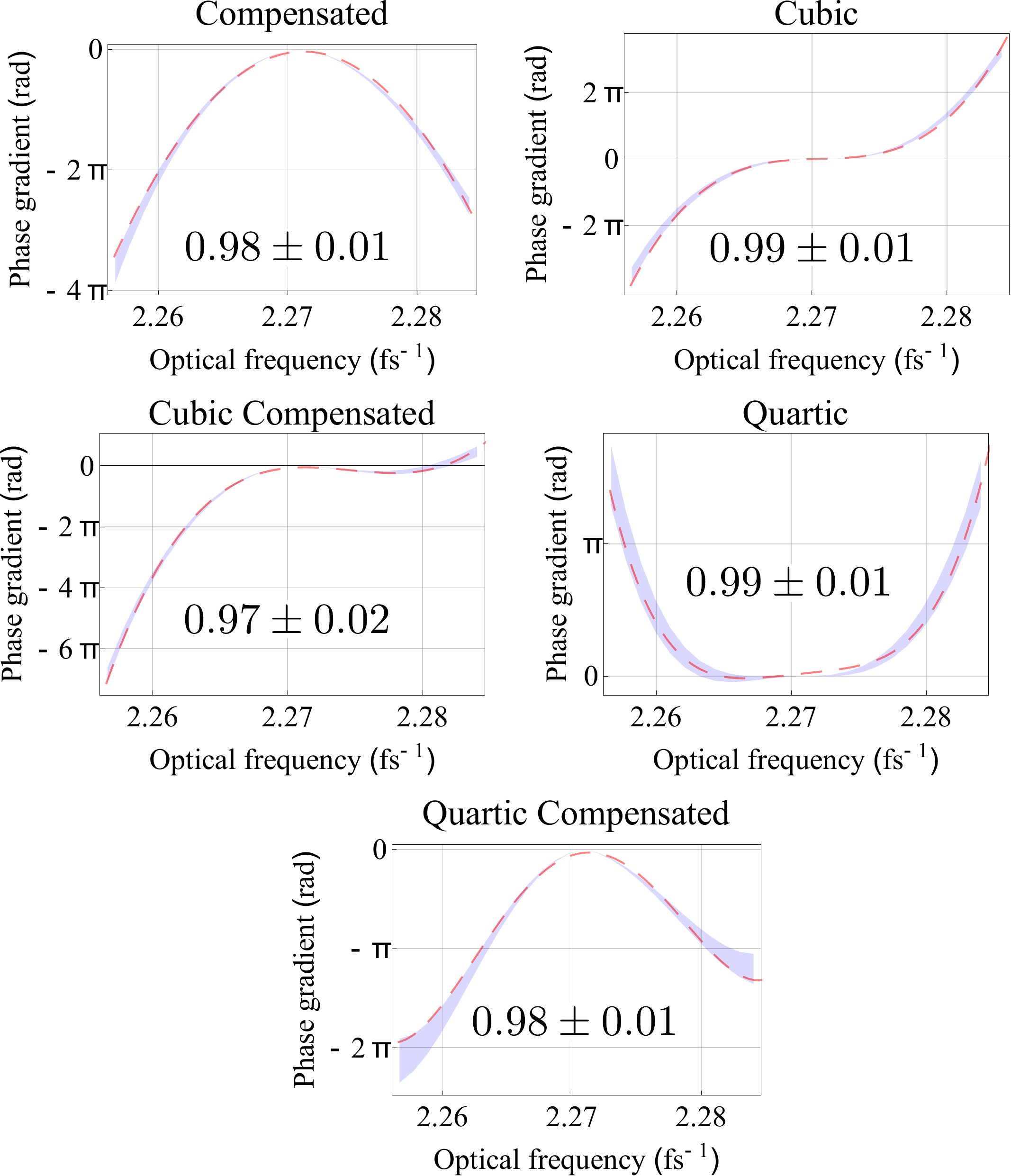}
	\caption{Comparison between the phase written on the pulse shaper (dashed) and the reconstructed phase (plain). The shaded area represents the standard deviation of the reconstruction, which is naturally larger the further from $\omega_0$, where the phase is constrained. The inset numbers represent the fidelity of the reconstruction.}
	\label{shaper-phases}
\end{figure}

To further validate the calibration of the setup, we can compare the phase reconstruction to the known phase that was imprinted on the pulse shaper. This is achieved by computing the relative phase between the shaped and unshaped pulse. The results depicted by Fig. \ref{shaper-phases} show a good agreement between the reconstruction and the original phase. The discrepancy can be attributed to the inability of the pulse shaper to write large polynomial coefficients, since the phase is too steep on the edge of the spectrum, thus introducing errors. Indeed, it can be seen that the reconstruction for quartic phase is more unstable than the other cases, since such a phase shows very steep phase gradient close to the wings of the spectrum. However, it can still be reconstructed with high accuracy even in the uncompensated case where a large quadratic component is mixed. It is worth noting that both algorithms described in Sec.\ref{sec:algo} can be used to reconstruct polynomial phases, giving similar results. This shows that the setup and the algorithm is indeed able to reconstruct higher order phase structures at the single photon level. 

	\subsection{Arbitrary phase reconstruction}

A final set of measurements were achieved to test the reconstruction capabilities of the setup when the phase that is imprinted on the single photons may not be expanded on a polynomial basis. As detailed in \cite{PRL:18}, two sets of arbitrary phases were chosen, which resulted in similar spectral and temporal intensities, but actually formed two orthogonal modes. They are defined by a linear spectral phase proportional to $\modulus{\w-\w_0}$, where the positive (negative) linear phase is labelled V-phase ($\Lambda$-phase). The effect of such a phase on the wavepacket is to delay or advance the long-wavelength part of the spectrum in an opposite direction to the short-wavelength part.

\begin{figure}[ht!]
	\includegraphics[width=1\linewidth]{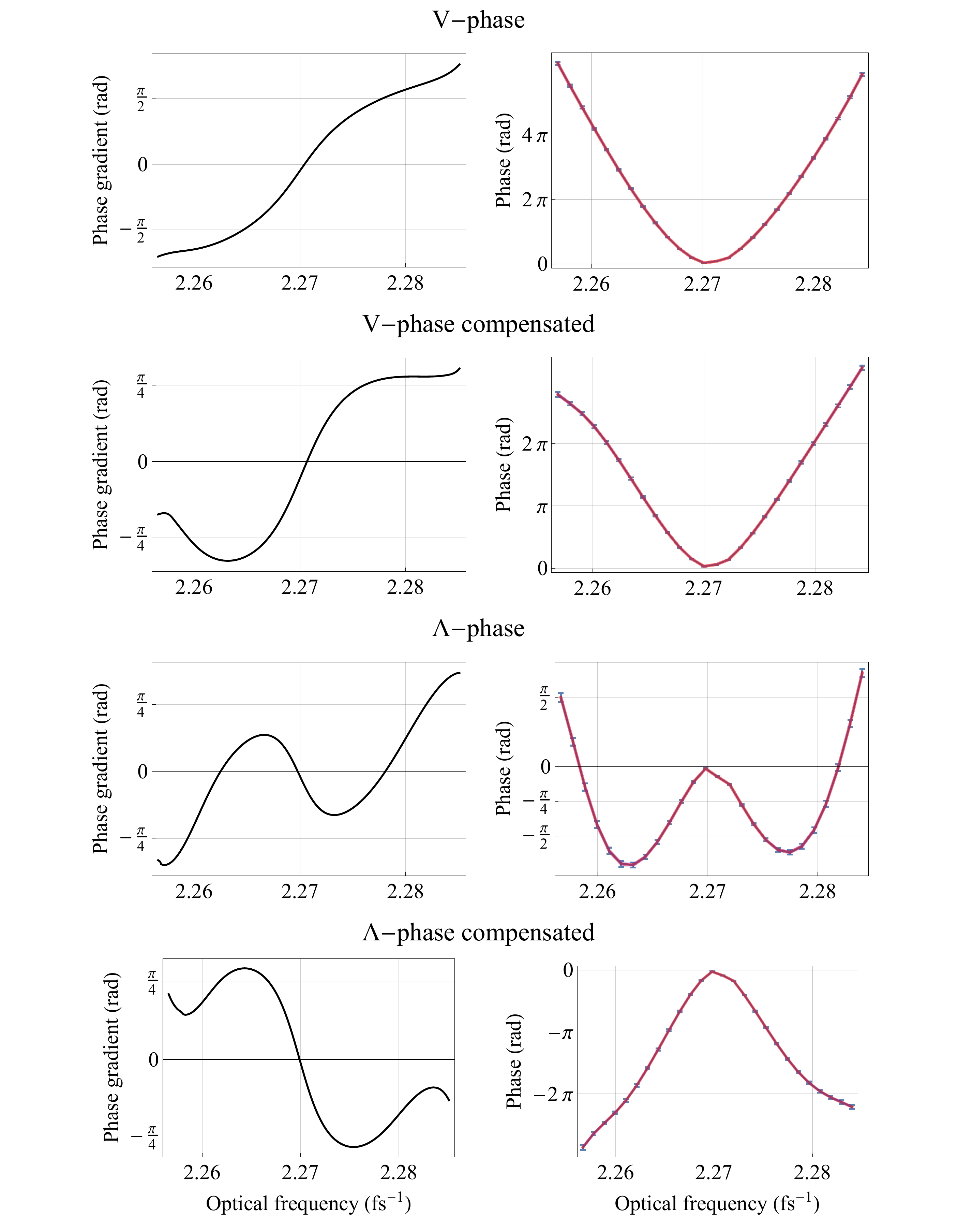}
	\caption{Left column: average of the extracted phase gradients corrected for delay; right column: reconstructed phases from the left column using the concatenation algorithm for a frequency shear $\W<0$. The error bars represent the standard deviation. The error is zero at the center frequency since it is constrained by the algorithm. The left column approximately shows the derivative of the right one.}
	\label{arb-phases}
\end{figure}

The value imprinted on the pulse shaper was $\pm 750$ fs, and a reconstruction was made. Creating phase profiles of this sort, which present a non-continuous spatial derivative of the phase on the mask of the SLM, can introduce diffraction, resulting in a small hole in the spectrum at the non-analytic point. Similarly to the previous section, these phases were reconstructed with and without compensation of the dispersion. The extracted and reconstructed phases are shown in Fig. \ref{arb-phases}. The extracted phase gradients $\Delta\phi(\omega,\Omega)$ closely resemble a step function where the steepness is determined by the value of the shear, and the value of the plateau reflect the amount of linear phase. Furthermore, we can clearly see an additional linear phase gradient in the uncompensated cases, showing again the presence of dispersion. The reconstructions show that the setup is able to easily distinguish the two phases. Moreover, a fit in the reconstruction gives an absolute linear phase of $+750 \pm 25$ fs for the V-phase and $-775 \pm 50$ fs for the $\Lambda$-phase, and the same quadratic phase found in the previous section for the uncompensated and compensated phases. The $\Lambda$-phase is more difficult to reconstruct compared to the previous phases due to the constraints put on the pulse shaper. Indeed, this phase has a steep negative slope, and compensating for dispersion adds an extra negative quadratic phase on the mask, resulting in additional wrappings.

	\subsection{Time-frequency distributions}
With the phase now reconstructed, it is possible to create time-frequency distributions to fully describe the spectral-temporal wave function of the single photon. Indeed, as it is the case with classical pulses, a complete description may be obtained by showing how the spectral parts are mapped in the time domain. Multiple ways to describe this distribution exist, and we choose the chronocyclic Wigner function:

\begin{align}
W(\omega,t) = \int \, \ud \theta \, \tilde{\varepsilon}^*(\omega-\theta/2) \, e^{-i \theta t} \, \tilde{\varepsilon}(\omega+\theta/2)
\label{eq:wigner}
\end{align}

\begin{figure}[t!]
	\includegraphics[width=1\linewidth]{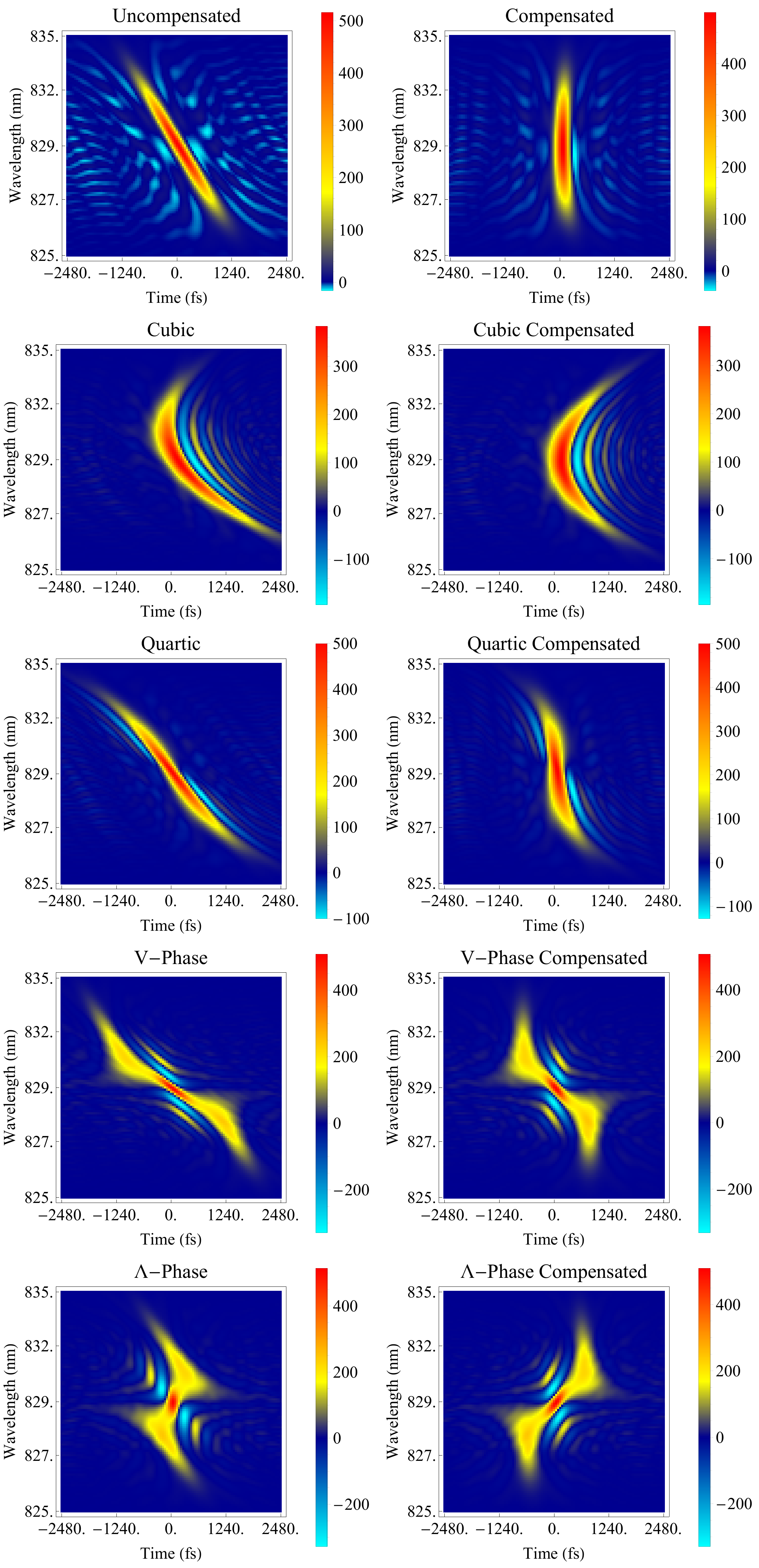}
	\caption{Chronocyclic Wigner function for the phases that were imprinted with the pulse shaper.}
	\label{all-wigner}
\end{figure}

The chronocyclic Wigner function has the advantages that it provides a complete characterization of the single-photon spectral-temporal density matrix, is normalized and real-valued everywhere, and the overlap of two states can be easily calculated from the overlap of the respective Wigner functions \cite{Paye:92}. It is often referred to as a ``quasiprobability distribution" since its marginals correspond to the probability distributions of measurement outcomes in those bases, although the function itself can locally take negative values. To compute such distribution, one needs the complex spectral amplitude $\modulus{\tilde{\varepsilon}(\w)}$ in order to reconstruct the complete spectral wave function with the phase that was measured. In principle this can be directly measured during the acquisition by summing the two outputs of the interferometer, which removes the interference fringes. The result is quite close to the true spectrum for small values of the shear. It is also possible to independently measure the amplitude directly without sending pulses into the interferometer. The latter method was used here. The center frequency of the spectrum (i.e. the first moment of the distribution) is used as the experimental value of $\w_0$ throughout the algorithm, notably to constrain the center frequency of the phase concatenation and the fits. The time-frequency distributions for the polynomial and the arbitrary phases written with the pulse shaper are shown on Fig. \ref{all-wigner}.

Summing the distribution along one dimension yields either the temporal or the spectral marginals. As expected for polynomial phase, the temporal wave function of the uncompensated single photon is much broader than that of the compensated one, cubic phase creates replicas, and quartic phase creates a pedestral. In the case of the V-phase and $\Lambda$-phase, both marginals are similar, yet the two modes are orthogonal. Computing the overlap between these modes yield a value of $0.06 \pm 0.01$, while the overlap of their marginal temporal intensities is $0.95 \pm 0.01$ and the spectral overlap is unity.

\section{Applicability and challenges}
	\subsection{Mixed state reconstruction}
	The methods developed thus far are aimed at reconstruction of the pulse-mode state of pure single-photon sources, i.e. single-photon sources that emit photons into only one pulse mode. For sources that emit photons into a mixture of pulsed modes, we must generalize the state reconstruction process to reconstruct the full density operator $\hat{\rho}$. The problem of characterizing a mixed state single-photon source is similar to characterization of a partially-coherent pulse train \cite{bourassin:15}. 
	The result of a measurement with the EOSI is defined as the quantum counterpart of Eq. \ref{eqIntensityPattern} as $I_\W^\pm(\w) = \operatorname{Tr}\left[ \hat{\rho}\hat{\Pi}_\pm(\w) \right]$, where $\hat \Pi_\pm(\w)$ is the positive-operator valued measure (POVM) that describes the measurement of a single photon in a mixed state using the experimental scheme. It is defined by
\begin{align}
\hat{\Pi}_\pm(\w) = &\frac{1}{2}\left( \ket{\w}\bra{\w} + \ket{\w+\W}\bra{\w+\W} \right) \nonumber\\
                &\pm \ket{\w}\bra{\w+\W}e^{i\w\tau} \pm \ket{\w+\W}\bra{\w}e^{-i\w\tau}
\end{align}
	For a pure state, it is straightforward to show that this POVM is the one associated to Eq. \ref{eqIntensityPattern}. For a mixed state, by defining $\w' = \w + \W$ and scanning the shear, the 2-dimensional intensity pattern is then given by
\begin{align}
I_\W^\pm(\w,\w') = &\frac{1}{2}\left( \rho(\w,\w) + \rho(\w',\w')  \right) \nonumber\\
&\pm \operatorname{Re}\left\{ \rho(\w,\w') e^{i\w\tau} \right\}
\end{align}
	Hence, by scanning the spectral shift $\W$, it is possible to measure the spectral density matrix $\rho(\w,\w+\W)$ using 2D-Fourier analysis similar to the one presented in Sec. \ref{background}. Moreover, it may be easier to use the signal associated to the difference of the photocurrents at the outputs of the interferometer $I_\W^\textrm{diff}(\w,\w') = I_\W^+(\w,\w') - I_\W^-(\w,\w') = 2\operatorname{Re}\left\{ \rho(\w,\w') e^{i\w\tau} \right\}$, which directly samples the off-diagonal entries of the spectral density matrix. Note that adding a $\pi/2$ phase shift in the interferometer yields $2\operatorname{Im}\left\{ \rho(\w,\w') e^{i\w\tau} \right\}$, which allows full reconstruction of the density matrix. Even though the full reconstruction would prove more difficult due to the increase in dimension and need to scan the spectral shift, it is still feasible with this method.
	
	It is worth noting that this only samples elements of the spectral density matrix that are within the shear of the main diagonal. Full characterisaiton of an unknown state would require scanning the frequency shift on a range that is on the order of the spectral bandwidth. With the capabilities outlined in this paper, a mixed state of narrower bandwidth could in principle be reconstructed.
	
	\subsection{Challenges}
	While the general scheme depicted by Fig. \ref{fig:schematic} is not particularly complex, the experimental implementation places some limitations on the input that EOSI can realistically characterise. Altogether, these constraints are in general much more lax for EOSI than those required by externally-referencing methods. Here, we detail these constraints, and discuss how to choose the proper settings for the experimental setup.
	
The device is based on a polarization Mach-Zehnder configuration, which requires an input with well-defined polarization. Polarization filtering may be employed to achieve this, but this incurs losses and precludes the device being used to explore polarization-time/frequency correlations.
	
	A main building block of the EOSI is a high-resolution single-photon spectrometer which we construct using frequency-to-time conversion. While modern detectors provide sufficient temporal resolution, the methods for obtaining large chirped pulses are still limited. It is not feasible to stretch pulses sufficiently in free space to enable frequency-to-time conversion, so a fiber setup is needed. In the near infrared, optical fibers show typical dispersion on the order of 100 ps/nm/km and propagation losses of more than 3 dB/km, rendering large chirps with fiber propagation largely impractical. CFBGs prove to be the most efficient way of achieving the dispersion needed by the setup, but typically act in reflection over a limited spectral region, which is mainly defined by the length of the fiber and the specified dispersion. Larger spectral windows and larger chirp parameters could in principle be obtained by concatenating CFBG together. The association of the CFBG parameters and the timing jitter of the detectors then limit both the spectral range and the spectral resolution\cite{davis:17}, thus requiring \emph{a priori} knowledge of the center wavelength and bandwidth of the unknown single photon state.
	
	The other main limitation comes from the EOM, with which the spectral shift is obtained. Various methods can be used to apply a high-power linear RF signal phase-locked to the pulse train. While we use frequency multiplication and a phase-lock loop, we propose it is possible to amplify the bandpass-filtered signal from a large bandwidth photodiode and use it as the RF-wave which is automatically locked to the pulse train. As described by Eq. \ref{linearregion}, the amplitude of the frequency shear is directly proportional to both the frequency and the amplitude of the RF wave. The amplitude is limited by the amplifiers and by the EOM itself, while the frequency is bounded by the temporal width of the linear region of the RF-wave. If it is too large, the temporal pulse shape will be changed and thus introduce errors into the reconstruction. Therefore the device input must be a train of single photons with well-defined arrival times and it cannot be applied to continuous-wave single-photon sources. Such a problem could be tackled by replacing the sinusoidal RF wave by a serrodyne wave \cite{johnson:88}, which would not limit the temporal aperture of the device. The need for a large frequency shear is obvious from Eq. \ref{powerseries}, since it provides a larger amplitude of the phase gradient. It also needs to be larger than the resolution of the spectrometer. This parameter is however also bounded by the unknown state, as any phase structure smaller than the shear cannot be reconstructed. One could, in principle, concatenate several modulators (or multi-pass a single modulator) to increase the shear. However, this results in further insertion loss, timing jitter and dispersion that will adversely affect the calibration.

\section{Conclusions}
In summary, we have demonstrated a self-referencing technique to reconstruct the spectral-temporal component of a pulsed single photon wave function. This method is applicable across a wide range of wavelengths and pulse characteristics, and is deterministic in that only technical losses prevent every photon entering the device contributing to the final reconstruction. The method can be readily extended to the characterization of mixed state single-photon sources, as well as multi-photon states. In its present design our device has wide applications in single-photon source and detector characterization and quantum metrology. We anticipate that future experiments will utilize this technology for more diverse purposes such as spectral-temporal entanglement characterization and state purity determination.  

\begin{acknowledgments}
	We are grateful to C. Dorrer, C. Radzewicz, M. G. Raymer, and I. A. Walmsley for fruitful discussions. This project has received funding from the European Union's Horizon 2020 research and innovation programme under Grant Agreement No. 665148, the United Kingdom Defense Science and Technology Laboratory (DSTL) under contract No. DSTLX-100092545, and the National Science Foundation under Grant No. 1620822. This work was partially funded by the HOMING programme of the Foundation for Polish Science (project no. Homing/2016-1/4) co-financed by the European Union under the European Regional Development Fund.
\end{acknowledgments}

\section*{References}
\bibliography{bibliography}

\begin{thebibliography}{36}%
\makeatletter
\providecommand \@ifxundefined [1]{%
 \@ifx{#1\undefined}
}%
\providecommand \@ifnum [1]{%
 \ifnum #1\expandafter \@firstoftwo
 \else \expandafter \@secondoftwo
 \fi
}%
\providecommand \@ifx [1]{%
 \ifx #1\expandafter \@firstoftwo
 \else \expandafter \@secondoftwo
 \fi
}%
\providecommand \natexlab [1]{#1}%
\providecommand \enquote  [1]{``#1''}%
\providecommand \bibnamefont  [1]{#1}%
\providecommand \bibfnamefont [1]{#1}%
\providecommand \citenamefont [1]{#1}%
\providecommand \href@noop [0]{\@secondoftwo}%
\providecommand \href [0]{\begingroup \@sanitize@url \@href}%
\providecommand \@href[1]{\@@startlink{#1}\@@href}%
\providecommand \@@href[1]{\endgroup#1\@@endlink}%
\providecommand \@sanitize@url [0]{\catcode `\\12\catcode `\$12\catcode
  `\&12\catcode `\#12\catcode `\^12\catcode `\_12\catcode `\%12\relax}%
\providecommand \@@startlink[1]{}%
\providecommand \@@endlink[0]{}%
\providecommand \url  [0]{\begingroup\@sanitize@url \@url }%
\providecommand \@url [1]{\endgroup\@href {#1}{\urlprefix }}%
\providecommand \urlprefix  [0]{URL }%
\providecommand \Eprint [0]{\href }%
\providecommand \doibase [0]{http://dx.doi.org/}%
\providecommand \selectlanguage [0]{\@gobble}%
\providecommand \bibinfo  [0]{\@secondoftwo}%
\providecommand \bibfield  [0]{\@secondoftwo}%
\providecommand \translation [1]{[#1]}%
\providecommand \BibitemOpen [0]{}%
\providecommand \bibitemStop [0]{}%
\providecommand \bibitemNoStop [0]{.\EOS\space}%
\providecommand \EOS [0]{\spacefactor3000\relax}%
\providecommand \BibitemShut  [1]{\csname bibitem#1\endcsname}%
\let\auto@bib@innerbib\@empty
%</preamble>
\bibitem [{\citenamefont {Humphreys}\ \emph {et~al.}(2014)\citenamefont
  {Humphreys} \emph {et~al.}}]{humphreys:14}%
  \BibitemOpen
  \bibfield  {author} {\bibinfo {author} {\bibfnamefont {P.}~\bibnamefont
  {Humphreys}} \emph {et~al.},\ }\href@noop {} {\bibfield  {journal} {\bibinfo
  {journal} {Phys. Rev. Lett.}\ }\textbf {\bibinfo {volume} {113}},\ \bibinfo
  {pages} {130502} (\bibinfo {year} {2014})}\BibitemShut {NoStop}%
\bibitem [{\citenamefont {Lamine}\ \emph {et~al.}(2008)\citenamefont {Lamine},
  \citenamefont {Fabre},\ and\ \citenamefont {Treps}}]{lamine:08}%
  \BibitemOpen
  \bibfield  {author} {\bibinfo {author} {\bibfnamefont {B.}~\bibnamefont
  {Lamine}}, \bibinfo {author} {\bibfnamefont {C.}~\bibnamefont {Fabre}}, \
  and\ \bibinfo {author} {\bibfnamefont {N.}~\bibnamefont {Treps}},\
  }\href@noop {} {\bibfield  {journal} {\bibinfo  {journal} {Physical Review
  Letters}\ }\textbf {\bibinfo {volume} {101}},\ \bibinfo {pages} {123601}
  (\bibinfo {year} {2008})}\BibitemShut {NoStop}%
\bibitem [{\citenamefont {Jian}\ \emph {et~al.}(2012)\citenamefont {Jian},
  \citenamefont {Pinel}, \citenamefont {Fabre}, \citenamefont {Lamine},\ and\
  \citenamefont {Treps}}]{jian2012real}%
  \BibitemOpen
  \bibfield  {author} {\bibinfo {author} {\bibfnamefont {P.}~\bibnamefont
  {Jian}}, \bibinfo {author} {\bibfnamefont {O.}~\bibnamefont {Pinel}},
  \bibinfo {author} {\bibfnamefont {C.}~\bibnamefont {Fabre}}, \bibinfo
  {author} {\bibfnamefont {B.}~\bibnamefont {Lamine}}, \ and\ \bibinfo {author}
  {\bibfnamefont {N.}~\bibnamefont {Treps}},\ }\href@noop {} {\bibfield
  {journal} {\bibinfo  {journal} {Optics express}\ }\textbf {\bibinfo {volume}
  {20}},\ \bibinfo {pages} {27133} (\bibinfo {year} {2012})}\BibitemShut
  {NoStop}%
\bibitem [{\citenamefont {Nunn}\ \emph {et~al.}(2013)\citenamefont {Nunn} \emph
  {et~al.}}]{nunn:13}%
  \BibitemOpen
  \bibfield  {author} {\bibinfo {author} {\bibfnamefont {J.}~\bibnamefont
  {Nunn}} \emph {et~al.},\ }\href@noop {} {\bibfield  {journal} {\bibinfo
  {journal} {Optics Express}\ }\textbf {\bibinfo {volume} {21}},\ \bibinfo
  {pages} {15959} (\bibinfo {year} {2013})}\BibitemShut {NoStop}%
\bibitem [{\citenamefont {Lounis}\ and\ \citenamefont
  {Orrit}(2005)}]{lounis:05}%
  \BibitemOpen
  \bibfield  {author} {\bibinfo {author} {\bibfnamefont {B.}~\bibnamefont
  {Lounis}}\ and\ \bibinfo {author} {\bibfnamefont {M.}~\bibnamefont {Orrit}},\
  }\href@noop {} {\bibfield  {journal} {\bibinfo  {journal} {Reports on
  Progress in Physics}\ }\textbf {\bibinfo {volume} {68}},\ \bibinfo {pages}
  {1129} (\bibinfo {year} {2005})}\BibitemShut {NoStop}%
\bibitem [{\citenamefont {Qin}\ \emph {et~al.}(2015)\citenamefont {Qin},
  \citenamefont {Prasad}, \citenamefont {Brannan}, \citenamefont {MacRae},
  \citenamefont {Lezama},\ and\ \citenamefont {Lvovsky}}]{qin:15}%
  \BibitemOpen
  \bibfield  {author} {\bibinfo {author} {\bibfnamefont {Z.}~\bibnamefont
  {Qin}}, \bibinfo {author} {\bibfnamefont {A.}~\bibnamefont {Prasad}},
  \bibinfo {author} {\bibfnamefont {T.}~\bibnamefont {Brannan}}, \bibinfo
  {author} {\bibfnamefont {A.}~\bibnamefont {MacRae}}, \bibinfo {author}
  {\bibfnamefont {A.}~\bibnamefont {Lezama}}, \ and\ \bibinfo {author}
  {\bibfnamefont {A.}~\bibnamefont {Lvovsky}},\ }\href@noop {} {\bibfield
  {journal} {\bibinfo  {journal} {Light: Science and Applications}\ }\textbf
  {\bibinfo {volume} {4}} (\bibinfo {year} {2015})}\BibitemShut {NoStop}%
\bibitem [{\citenamefont {Weiner}(2011{\natexlab{a}})}]{weiner2011Book}%
  \BibitemOpen
  \bibfield  {author} {\bibinfo {author} {\bibfnamefont {A.}~\bibnamefont
  {Weiner}},\ }\href@noop {} {\emph {\bibinfo {title} {Ultrafast optics}}},\
  Vol.~\bibinfo {volume} {72}\ (\bibinfo  {publisher} {John Wiley \& Sons},\
  \bibinfo {year} {2011})\BibitemShut {NoStop}%
\bibitem [{\citenamefont {Walmsley}\ and\ \citenamefont
  {Dorrer}(2009)}]{walmsley:09}%
  \BibitemOpen
  \bibfield  {author} {\bibinfo {author} {\bibfnamefont {I.}~\bibnamefont
  {Walmsley}}\ and\ \bibinfo {author} {\bibfnamefont {C.}~\bibnamefont
  {Dorrer}},\ }\href@noop {} {\bibfield  {journal} {\bibinfo  {journal}
  {Advances in Optics and Photonics}\ }\textbf {\bibinfo {volume} {1}},\
  \bibinfo {pages} {308} (\bibinfo {year} {2009})}\BibitemShut {NoStop}%
\bibitem [{\citenamefont {Wasilewski}\ \emph {et~al.}(2007)\citenamefont
  {Wasilewski}, \citenamefont {Kolenderski},\ and\ \citenamefont
  {Frankowski}}]{wasilewski:07}%
  \BibitemOpen
  \bibfield  {author} {\bibinfo {author} {\bibfnamefont {W.}~\bibnamefont
  {Wasilewski}}, \bibinfo {author} {\bibfnamefont {P.}~\bibnamefont
  {Kolenderski}}, \ and\ \bibinfo {author} {\bibfnamefont {R.}~\bibnamefont
  {Frankowski}},\ }\href@noop {} {\bibfield  {journal} {\bibinfo  {journal}
  {Physical Review Letters}\ }\textbf {\bibinfo {volume} {99}},\ \bibinfo
  {pages} {123601} (\bibinfo {year} {2007})}\BibitemShut {NoStop}%
\bibitem [{\citenamefont {Brecht}\ \emph {et~al.}(2015)\citenamefont {Brecht},
  \citenamefont {Reddy}, \citenamefont {Silberhorn},\ and\ \citenamefont
  {Raymer}}]{brecht:15}%
  \BibitemOpen
  \bibfield  {author} {\bibinfo {author} {\bibfnamefont {B.}~\bibnamefont
  {Brecht}}, \bibinfo {author} {\bibfnamefont {D.~V.}\ \bibnamefont {Reddy}},
  \bibinfo {author} {\bibfnamefont {C.}~\bibnamefont {Silberhorn}}, \ and\
  \bibinfo {author} {\bibfnamefont {M.~G.}\ \bibnamefont {Raymer}},\
  }\href@noop {} {\bibfield  {journal} {\bibinfo  {journal} {Phys. Rev. X}\
  }\textbf {\bibinfo {volume} {5}},\ \bibinfo {pages} {041017} (\bibinfo {year}
  {2015})}\BibitemShut {NoStop}%
\bibitem [{\citenamefont {Polycarpou}\ \emph {et~al.}(2012)\citenamefont
  {Polycarpou}, \citenamefont {Cassemiro}, \citenamefont {Venturi},
  \citenamefont {Zavatta},\ and\ \citenamefont {Bellini}}]{polycarpou:12}%
  \BibitemOpen
  \bibfield  {author} {\bibinfo {author} {\bibfnamefont {C.}~\bibnamefont
  {Polycarpou}}, \bibinfo {author} {\bibfnamefont {K.~N.}\ \bibnamefont
  {Cassemiro}}, \bibinfo {author} {\bibfnamefont {G.}~\bibnamefont {Venturi}},
  \bibinfo {author} {\bibfnamefont {A.}~\bibnamefont {Zavatta}}, \ and\
  \bibinfo {author} {\bibfnamefont {M.}~\bibnamefont {Bellini}},\ }\href@noop
  {} {\bibfield  {journal} {\bibinfo  {journal} {Physical Review Letters}\
  }\textbf {\bibinfo {volume} {109}},\ \bibinfo {pages} {053602} (\bibinfo
  {year} {2012})}\BibitemShut {NoStop}%
\bibitem [{\citenamefont {Trebino}\ \emph {et~al.}(1997)\citenamefont {Trebino}
  \emph {et~al.}}]{trebino:97}%
  \BibitemOpen
  \bibfield  {author} {\bibinfo {author} {\bibfnamefont {R.}~\bibnamefont
  {Trebino}} \emph {et~al.},\ }\href@noop {} {\bibfield  {journal} {\bibinfo
  {journal} {Rev. Sci. Instrum}\ }\textbf {\bibinfo {volume} {68}} (\bibinfo
  {year} {1997})}\BibitemShut {NoStop}%
\bibitem [{\citenamefont {Brecht}\ \emph {et~al.}(2011)\citenamefont {Brecht}
  \emph {et~al.}}]{brecht:11}%
  \BibitemOpen
  \bibfield  {author} {\bibinfo {author} {\bibfnamefont {B.}~\bibnamefont
  {Brecht}} \emph {et~al.},\ }\href@noop {} {\bibfield  {journal} {\bibinfo
  {journal} {New Journal of Physics}\ }\textbf {\bibinfo {volume} {13}},\
  \bibinfo {pages} {065029} (\bibinfo {year} {2011})}\BibitemShut {NoStop}%
\bibitem [{\citenamefont {Dorrer}\ and\ \citenamefont
  {Kang}(2003)}]{dorrer:03}%
  \BibitemOpen
  \bibfield  {author} {\bibinfo {author} {\bibfnamefont {C.}~\bibnamefont
  {Dorrer}}\ and\ \bibinfo {author} {\bibfnamefont {I.}~\bibnamefont {Kang}},\
  }\href@noop {} {\bibfield  {journal} {\bibinfo  {journal} {Optics Letters}\
  }\textbf {\bibinfo {volume} {28}},\ \bibinfo {pages} {477} (\bibinfo {year}
  {2003})}\BibitemShut {NoStop}%
\bibitem [{\citenamefont {Wright}\ \emph {et~al.}(2017)\citenamefont {Wright}
  \emph {et~al.}}]{wright:17}%
  \BibitemOpen
  \bibfield  {author} {\bibinfo {author} {\bibfnamefont {L.}~\bibnamefont
  {Wright}} \emph {et~al.},\ }\href@noop {} {\bibfield  {journal} {\bibinfo
  {journal} {Physical Review Letters}\ }\textbf {\bibinfo {volume} {118}},\
  \bibinfo {pages} {023601} (\bibinfo {year} {2017})}\BibitemShut {NoStop}%
\bibitem [{\citenamefont {Davis}\ \emph {et~al.}(2018)\citenamefont {Davis},
  \citenamefont {Thiel}, \citenamefont {Karpi\'{n}ski},\ and\ \citenamefont
  {Smith}}]{PRL:18}%
  \BibitemOpen
  \bibfield  {author} {\bibinfo {author} {\bibfnamefont {A.}~\bibnamefont
  {Davis}}, \bibinfo {author} {\bibfnamefont {V.}~\bibnamefont {Thiel}},
  \bibinfo {author} {\bibfnamefont {M.}~\bibnamefont {Karpi\'{n}ski}}, \ and\
  \bibinfo {author} {\bibfnamefont {B.}~\bibnamefont {Smith}},\ }\href@noop {}
  {\bibfield  {journal} {\bibinfo  {journal} {In preparation}\ } (\bibinfo
  {year} {2018})}\BibitemShut {NoStop}%
\bibitem [{\citenamefont {Bialynicki-Birula}(1996)}]{birula:96}%
  \BibitemOpen
  \bibfield  {author} {\bibinfo {author} {\bibfnamefont {I.}~\bibnamefont
  {Bialynicki-Birula}},\ }in\ \href@noop {} {\emph {\bibinfo {booktitle}
  {Progress in Optics}}},\ Vol.~\bibinfo {volume} {36}\ (\bibinfo  {publisher}
  {Elsevier},\ \bibinfo {year} {1996})\ pp.\ \bibinfo {pages}
  {245--294}\BibitemShut {NoStop}%
\bibitem [{\citenamefont {Sipe}(1995)}]{sipe:95}%
  \BibitemOpen
  \bibfield  {author} {\bibinfo {author} {\bibfnamefont {J.}~\bibnamefont
  {Sipe}},\ }\href@noop {} {\bibfield  {journal} {\bibinfo  {journal} {Physical
  Review A}\ }\textbf {\bibinfo {volume} {52}},\ \bibinfo {pages} {1875}
  (\bibinfo {year} {1995})}\BibitemShut {NoStop}%
\bibitem [{\citenamefont {Smith}\ and\ \citenamefont
  {Raymer}(2007)}]{smith:07}%
  \BibitemOpen
  \bibfield  {author} {\bibinfo {author} {\bibfnamefont {B.}~\bibnamefont
  {Smith}}\ and\ \bibinfo {author} {\bibfnamefont {M.}~\bibnamefont {Raymer}},\
  }\href@noop {} {\bibfield  {journal} {\bibinfo  {journal} {New Journal of
  Physics}\ }\textbf {\bibinfo {volume} {9}} (\bibinfo {year}
  {2007})}\BibitemShut {NoStop}%
\bibitem [{\citenamefont {Dorrer}\ \emph {et~al.}(2002)\citenamefont {Dorrer},
  \citenamefont {Kosik},\ and\ \citenamefont {Walmsley}}]{kosik:02}%
  \BibitemOpen
  \bibfield  {author} {\bibinfo {author} {\bibfnamefont {C.}~\bibnamefont
  {Dorrer}}, \bibinfo {author} {\bibfnamefont {E.}~\bibnamefont {Kosik}}, \
  and\ \bibinfo {author} {\bibfnamefont {I.}~\bibnamefont {Walmsley}},\
  }\href@noop {} {\bibfield  {journal} {\bibinfo  {journal} {Optics Letters}\
  }\textbf {\bibinfo {volume} {27}},\ \bibinfo {pages} {548} (\bibinfo {year}
  {2002})}\BibitemShut {NoStop}%
\bibitem [{\citenamefont {Dorrer}\ \emph {et~al.}(2014)\citenamefont {Dorrer},
  \citenamefont {Kosik},\ and\ \citenamefont {Walmsley}}]{kosik:14}%
  \BibitemOpen
  \bibfield  {author} {\bibinfo {author} {\bibfnamefont {C.}~\bibnamefont
  {Dorrer}}, \bibinfo {author} {\bibfnamefont {E.}~\bibnamefont {Kosik}}, \
  and\ \bibinfo {author} {\bibfnamefont {I.}~\bibnamefont {Walmsley}},\
  }\href@noop {} {\bibfield  {journal} {\bibinfo  {journal} {Applied Physics
  B.}\ }\textbf {\bibinfo {volume} {74}},\ \bibinfo {pages} {s209} (\bibinfo
  {year} {2014})}\BibitemShut {NoStop}%
\bibitem [{\citenamefont {Dorrer}\ and\ \citenamefont
  {Walmsley}(2001)}]{dorrer:01}%
  \BibitemOpen
  \bibfield  {author} {\bibinfo {author} {\bibfnamefont {C.}~\bibnamefont
  {Dorrer}}\ and\ \bibinfo {author} {\bibfnamefont {I.}~\bibnamefont
  {Walmsley}},\ }\href@noop {} {\bibfield  {journal} {\bibinfo  {journal}
  {Journal of the Optical Society of America B}\ }\textbf {\bibinfo {volume}
  {19}},\ \bibinfo {pages} {1019} (\bibinfo {year} {2001})}\BibitemShut
  {NoStop}%
\bibitem [{\citenamefont {Dorrer}\ and\ \citenamefont
  {Bromage}(2010)}]{dorrer:10}%
  \BibitemOpen
  \bibfield  {author} {\bibinfo {author} {\bibfnamefont {C.}~\bibnamefont
  {Dorrer}}\ and\ \bibinfo {author} {\bibfnamefont {J.}~\bibnamefont
  {Bromage}},\ }\href@noop {} {\bibfield  {journal} {\bibinfo  {journal}
  {Optics Letters}\ }\textbf {\bibinfo {volume} {35}} (\bibinfo {year}
  {2010})}\BibitemShut {NoStop}%
\bibitem [{\citenamefont {Mosley}\ \emph {et~al.}(2008)\citenamefont {Mosley},
  \citenamefont {Lundeen}, \citenamefont {Smith}, \citenamefont {Wasylczyk},
  \citenamefont {U'Ren}, \citenamefont {Silberhorn},\ and\ \citenamefont
  {Walmsley}}]{mosley:08}%
  \BibitemOpen
  \bibfield  {author} {\bibinfo {author} {\bibfnamefont {P.~J.}\ \bibnamefont
  {Mosley}}, \bibinfo {author} {\bibfnamefont {J.~S.}\ \bibnamefont {Lundeen}},
  \bibinfo {author} {\bibfnamefont {B.~J.}\ \bibnamefont {Smith}}, \bibinfo
  {author} {\bibfnamefont {P.}~\bibnamefont {Wasylczyk}}, \bibinfo {author}
  {\bibfnamefont {A.~B.}\ \bibnamefont {U'Ren}}, \bibinfo {author}
  {\bibfnamefont {C.}~\bibnamefont {Silberhorn}}, \ and\ \bibinfo {author}
  {\bibfnamefont {I.~A.}\ \bibnamefont {Walmsley}},\ }\href@noop {} {\bibfield
  {journal} {\bibinfo  {journal} {Physical Review Letters}\ }\textbf {\bibinfo
  {volume} {100}},\ \bibinfo {pages} {133601} (\bibinfo {year}
  {2008})}\BibitemShut {NoStop}%
\bibitem [{\citenamefont {Moseley}\ \emph {et~al.}(2008)\citenamefont {Moseley}
  \emph {et~al.}}]{mosley:08b}%
  \BibitemOpen
  \bibfield  {author} {\bibinfo {author} {\bibfnamefont {P.}~\bibnamefont
  {Moseley}} \emph {et~al.},\ }\href@noop {} {\bibfield  {journal} {\bibinfo
  {journal} {New J. Phys.}\ }\textbf {\bibinfo {volume} {10}},\ \bibinfo
  {pages} {093011} (\bibinfo {year} {2008})}\BibitemShut {NoStop}%
\bibitem [{\citenamefont {Weiner}(2000)}]{weiner:00}%
  \BibitemOpen
  \bibfield  {author} {\bibinfo {author} {\bibfnamefont {A.~M.}\ \bibnamefont
  {Weiner}},\ }\href@noop {} {\bibfield  {journal} {\bibinfo  {journal} {Review
  of scientific instruments}\ }\textbf {\bibinfo {volume} {71}},\ \bibinfo
  {pages} {1929} (\bibinfo {year} {2000})}\BibitemShut {NoStop}%
\bibitem [{\citenamefont {Weiner}(2011{\natexlab{b}})}]{Weiner2011}%
  \BibitemOpen
  \bibfield  {author} {\bibinfo {author} {\bibfnamefont {A.~M.}\ \bibnamefont
  {Weiner}},\ }\href {\doibase 10.1016/j.optcom.2011.03.084} {\bibfield
  {journal} {\bibinfo  {journal} {Optics Communications}\ }\textbf {\bibinfo
  {volume} {284}},\ \bibinfo {pages} {3669} (\bibinfo {year}
  {2011}{\natexlab{b}})}\BibitemShut {NoStop}%
\bibitem [{\citenamefont {Karpi\'{n}ski}\ \emph {et~al.}(2017)\citenamefont
  {Karpi\'{n}ski} \emph {et~al.}}]{karpinski:17}%
  \BibitemOpen
  \bibfield  {author} {\bibinfo {author} {\bibfnamefont {M.}~\bibnamefont
  {Karpi\'{n}ski}} \emph {et~al.},\ }\href@noop {} {\bibfield  {journal}
  {\bibinfo  {journal} {Nature Photonics}\ }\textbf {\bibinfo {volume} {11}},\
  \bibinfo {pages} {55} (\bibinfo {year} {2017})}\BibitemShut {NoStop}%
\bibitem [{\citenamefont {Dorrer}(2008)}]{dorrer:08}%
  \BibitemOpen
  \bibfield  {author} {\bibinfo {author} {\bibfnamefont {C.}~\bibnamefont
  {Dorrer}},\ }\href@noop {} {\bibfield  {journal} {\bibinfo  {journal} {Optics
  Express}\ }\textbf {\bibinfo {volume} {16}},\ \bibinfo {pages} {6567}
  (\bibinfo {year} {2008})}\BibitemShut {NoStop}%
\bibitem [{\citenamefont {Avenhaus}\ \emph {et~al.}(2010)\citenamefont
  {Avenhaus} \emph {et~al.}}]{avenhaus:10}%
  \BibitemOpen
  \bibfield  {author} {\bibinfo {author} {\bibfnamefont {M.}~\bibnamefont
  {Avenhaus}} \emph {et~al.},\ }\href@noop {} {\bibfield  {journal} {\bibinfo
  {journal} {Physical Review Letters}\ }\textbf {\bibinfo {volume} {104}}
  (\bibinfo {year} {2010})}\BibitemShut {NoStop}%
\bibitem [{\citenamefont {Muriel}\ \emph {et~al.}(1999)\citenamefont {Muriel},
  \citenamefont {Aza{\~n}a},\ and\ \citenamefont {Carballar}}]{muriel1999}%
  \BibitemOpen
  \bibfield  {author} {\bibinfo {author} {\bibfnamefont {M.~A.}\ \bibnamefont
  {Muriel}}, \bibinfo {author} {\bibfnamefont {J.}~\bibnamefont {Aza{\~n}a}}, \
  and\ \bibinfo {author} {\bibfnamefont {A.}~\bibnamefont {Carballar}},\
  }\href@noop {} {\bibfield  {journal} {\bibinfo  {journal} {Optics letters}\
  }\textbf {\bibinfo {volume} {24}},\ \bibinfo {pages} {1} (\bibinfo {year}
  {1999})}\BibitemShut {NoStop}%
\bibitem [{\citenamefont {Aza{\~n}a}\ and\ \citenamefont
  {Muriel}(2000)}]{azana2000}%
  \BibitemOpen
  \bibfield  {author} {\bibinfo {author} {\bibfnamefont {J.}~\bibnamefont
  {Aza{\~n}a}}\ and\ \bibinfo {author} {\bibfnamefont {M.~A.}\ \bibnamefont
  {Muriel}},\ }\href@noop {} {\bibfield  {journal} {\bibinfo  {journal} {IEEE
  Journal of quantum electronics}\ }\textbf {\bibinfo {volume} {36}},\ \bibinfo
  {pages} {517} (\bibinfo {year} {2000})}\BibitemShut {NoStop}%
\bibitem [{\citenamefont {Davis}\ \emph {et~al.}(2017)\citenamefont {Davis},
  \citenamefont {Saulnier}, \citenamefont {Karpi\'{n}ski},\ and\ \citenamefont
  {Smith}}]{davis:17}%
  \BibitemOpen
  \bibfield  {author} {\bibinfo {author} {\bibfnamefont {A.}~\bibnamefont
  {Davis}}, \bibinfo {author} {\bibfnamefont {P.}~\bibnamefont {Saulnier}},
  \bibinfo {author} {\bibfnamefont {M.}~\bibnamefont {Karpi\'{n}ski}}, \ and\
  \bibinfo {author} {\bibfnamefont {B.}~\bibnamefont {Smith}},\ }\href@noop {}
  {\bibfield  {journal} {\bibinfo  {journal} {Optics Express}\ }\textbf
  {\bibinfo {volume} {25}},\ \bibinfo {pages} {12804} (\bibinfo {year}
  {2017})}\BibitemShut {NoStop}%
\bibitem [{\citenamefont {{Paye}}(1992)}]{Paye:92}%
  \BibitemOpen
  \bibfield  {author} {\bibinfo {author} {\bibfnamefont {J.}~\bibnamefont
  {{Paye}}},\ }\href {\doibase 10.1109/3.159533} {\bibfield  {journal}
  {\bibinfo  {journal} {IEEE Journal of Quantum Electronics}\ }\textbf
  {\bibinfo {volume} {28}},\ \bibinfo {pages} {2262} (\bibinfo {year}
  {1992})}\BibitemShut {NoStop}%
\bibitem [{\citenamefont {Bourassin-Bouchet}\ and\ \citenamefont
  {Couprie}(2015)}]{bourassin:15}%
  \BibitemOpen
  \bibfield  {author} {\bibinfo {author} {\bibfnamefont {C.}~\bibnamefont
  {Bourassin-Bouchet}}\ and\ \bibinfo {author} {\bibfnamefont {M.-E.}\
  \bibnamefont {Couprie}},\ }\href@noop {} {\bibfield  {journal} {\bibinfo
  {journal} {Nature communications}\ }\textbf {\bibinfo {volume} {6}},\
  \bibinfo {pages} {6465} (\bibinfo {year} {2015})}\BibitemShut {NoStop}%
\bibitem [{\citenamefont {Johnson}\ and\ \citenamefont
  {Cox}(1988)}]{johnson:88}%
  \BibitemOpen
  \bibfield  {author} {\bibinfo {author} {\bibfnamefont {L.~M.}\ \bibnamefont
  {Johnson}}\ and\ \bibinfo {author} {\bibfnamefont {C.~H.}\ \bibnamefont
  {Cox}},\ }\href@noop {} {\bibfield  {journal} {\bibinfo  {journal} {Journal
  of lightwave technology}\ }\textbf {\bibinfo {volume} {6}},\ \bibinfo {pages}
  {109} (\bibinfo {year} {1988})}\BibitemShut {NoStop}%
\end{thebibliography}%

\end{document}